\documentclass[aoas,MSNbibl,nameyear,dvips]{arximspdf}
\usepackage{stfloats}
\usepackage{graphicx}
\usepackage{breakurl}

\doi{10.1214/10-AOAS398}
\volume{5}
\issue{1}
\pubyear{2011}
\firstpage{5}
\lastpage{44}

\makeatletter
\def\@bmisc[#1]{%
  \get@battribute{unstr}%
  \common@pub@types%
  \let\bauthor\bbl@bauthor%
  \let\bhowpublished\@firstofone%
  \def\borganization##1{{\bauthor@style ##1}}%
}

\fnbelowfloat
\newcommand{\iid}{\stackrel{\mathrm{i.i.d.}}{\sim}}
\newcommand{\argmin}{\operatorname{arg\min}}
\let\epsilon\varepsilon
\makeatother

\begin{document}
\begin{frontmatter}

\title{A statistical analysis of multiple temperature proxies: Are
reconstructions of surface temperatures over the last 1000 years reliable?\thanksref{T2}}

\runtitle{A statistical analysis of multiple temperature proxies}

\relateddoi{T2}{Discussed in \doi{10.1214/10-AOAS398I}, \doi{10.1214/10-AOAS398M},
\doi{10.1214/10-AOAS398C}, \doi{10.1214/10-AOAS398L}, \doi{10.1214/10-AOAS398G},
\doi{10.1214/10-AOAS398D}, \doi{10.1214/10-AOAS398H},
\doi{10.1214/10-AOAS398B},
\doi{10.1214/10-AOAS398K},
\doi{10.1214/10-AOAS398E},
\doi{10.1214/10-AOAS398F},
\doi{10.1214/10-AOAS398J},
\doi{10.1214/10-AOAS409}; rejoinder at \doi{10.1214/10-AOAS398REJ}.}

\begin{aug}
\author[A]{\fnms{Blakeley B.} \snm{McShane}\ead[label=e1]{b-mcshane@kellogg.northwestern.edu}%
\ead[label=u1,url]{http://www.blakemcshane.com}}
and
\author[B]{\fnms{Abraham J.} \snm{Wyner}\corref{}\ead[label=e2]{ajw@wharton.upenn.edu}%
\ead[label=u2,url]{http://www.adiwyner.com}}

\runauthor{B. B. McShane and A. J. Wyner}

\affiliation{Northwestern University and the University of Pennsylvania}

\address[A]{Kellogg School of Management \\
Northwestern University \\
Leverone Hall \\
2001 Sheridan Road \\
Evanston, Illinois 60208 \\
USA\\
\printead{e1}\\
\printead{u1}}

\address[B]{Department of Statistics\\
The Wharton School \\
University of Pennsylvania \\
400 Jon M. Huntsman Hall \\
3730 Walnut Street \\
Philadelphia, Pennsylvania 19104 \\
USA\\
\printead{e2}\\
\printead{u2}}
\end{aug}

% HISTORY:
\received{\smonth{1} \syear{2010}}
\revised{\smonth{8} \syear{2010}}

% ABSTRACT
%
\begin{abstract}
Predicting historic temperatures based on tree rings, ice cores, and
other natural proxies is a difficult endeavor. The relationship
between proxies and temperature is weak and the number of proxies is
far larger than the number of target data points. Furthermore, the
data contain complex spatial and temporal dependence structures which
are not easily captured with simple models.

In this paper, we assess the reliability of such reconstructions and
their statistical significance against various null models. We find
that the proxies do not predict temperature significantly better than
random series generated independently of temperature. Furthermore,
various model specifications that perform similarly at predicting
temperature produce extremely different historical backcasts. Finally,
the proxies seem unable to forecast the high levels of and sharp run-up
in temperature in the 1990s either in-sample or from contiguous holdout
blocks, thus casting doubt on their ability to predict such phenomena
if in fact they occurred several hundred years ago.

We propose our own reconstruction of Northern Hemisphere average annual
land temperature over the last millennium, assess its reliability, and
compare it to those from the climate science literature. Our model
provides a similar reconstruction but has much wider standard errors,
reflecting the weak signal and large uncertainty encountered in this
setting.
\end{abstract}

% KEYWORDS
%


\end{frontmatter}

%s1 ###
\section{Introduction}\label{intro}

Paleoclimatology is the study of climate and climate~chan\-ge over the
scale of the entire history of earth. A particular area of focus~is temperature.
Since reliable temperature records typically exist for
only the last 150 years or fewer, paleoclimatologists use measurements
from tree rings, ice sheets,  and other natural phenomena to
estimate past temperature. The key idea is to use various artifacts of
historical periods which were strongly influenced by temperature and
which survive to the present. For example, Antarctic ice cores contain
ancient bubbles of air which can be dated quite accurately. The
temperature of that air can be approximated by measuring the ratio of
major ions and isotopes of oxygen and hydrogen. Similarly, tree rings
measured from old growth forests can be dated to annual resolution, and
features can be extracted which are known to be related to
temperature.\looseness=1

The ``proxy record'' is comprised of these and many other types of
data, including boreholes, corals, speleothems, and lake sediments [see
\citet{Brad99} for detailed descriptions]. The basic statistical problem
is quite easy to explain. Scientists extract, scale, and calibrate the
data. Then, a training set consisting of the part of the proxy record
which overlaps the modern instrumental period (i.e., the past 150
years) is constructed and used to build a model. Finally, the model,
which maps the proxy record to a surface temperature, is used to
backcast or ``reconstruct'' historical temperatures.\looseness=1

This effort to reconstruct our planet's climate history has become
linked to the topic of Anthropogenic Global Warming (AGW). On the one
hand, this is peculiar since paleoclimatological reconstructions can
provide evidence only for the \textit{detection} of global warming and
even then they constitute only one such source of evidence. The
principal sources of evidence for the detection of global warming and
in particular the \textit{attribution} of it to anthropogenic factors come
from basic science as well as General Circulation Models (GCMs) that
have been tuned to data accumulated during the instrumental period [\citet{IPCC07}].
These models show that carbon dioxide, when released into
the atmosphere in sufficient concentration, can force temperature increases.

\begin{figure}

\includegraphics{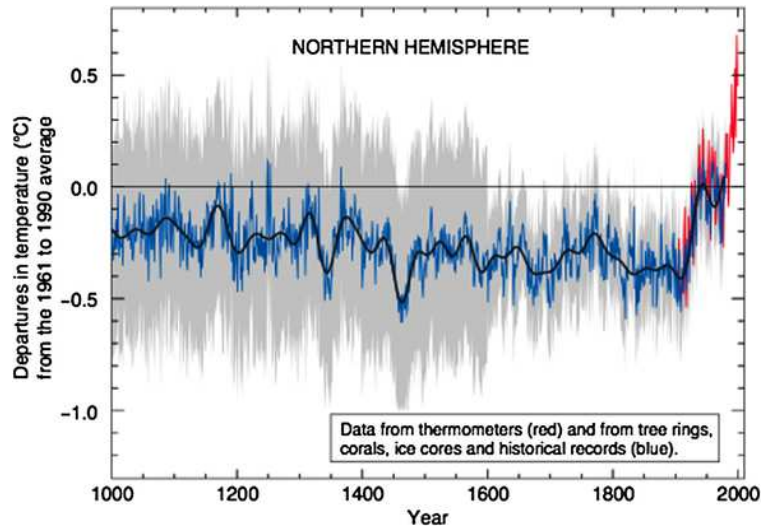}

\caption{Multiproxy reconstruction of Northern Hemisphere surface
temperature variations over the past millennium (blue), along with
$40$-year average (black), a measure of the statistical uncertainty
associated with the reconstruction (gray), and instrumental surface
temperature  (red), based on the work
by Mann, Bradley and Hughes (\protect\citeyear{MaBrHu99}). This figure has sometimes been referred
to as the ``hockey stick.'' Source: IPCC (\protect\citeyear{IPCC01}).}\label{fig:ipcc}
\end{figure}

\begin{figure}

\includegraphics{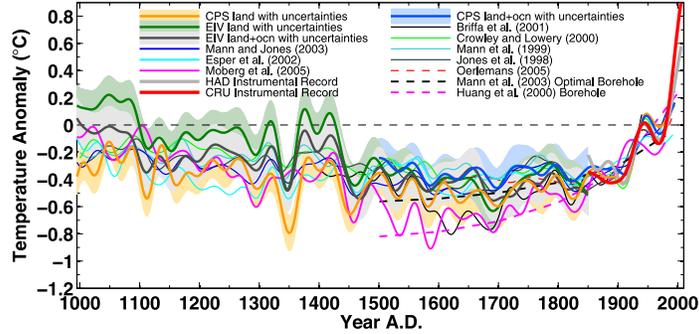}

\caption{Various reconstructions of Northern Hemisphere temperatures
over the last $1000$ years with $95$\% confidence intervals. Source:
Mann et~al. (\protect\citeyear{Mannetal08}).}\label{fig:thousand}
\end{figure}

\begin{figure}[b]

\includegraphics{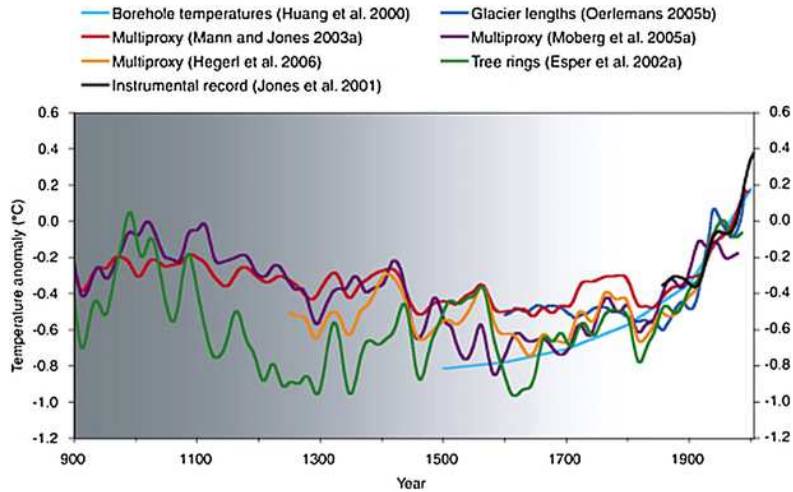}

\caption{Smoothed reconstructions of large-scale (Northern Hemisphere
mean or global mean) surface temperature variations from six
different research teams are shown along with the instrumental record
of global mean surface temperature. Each curve portrays a somewhat
different history of temperature variations and is subject to a
somewhat different set of uncertainties that generally increase going
backward in time (as indicated by the gray shading). Source: NRC (\protect\citeyear{NRC06}).}
\label{fig:nrc}
\vspace{-5pt}
\end{figure}

On the other hand, the effort of world governments to pass legislation
to cut carbon to pre-industrial levels cannot proceed without the
consent of the governed and historical reconstructions from
paleoclimatological models have indeed proven persuasive and effective
at winning the hearts and minds of the populace. %to impose regulations
%and spend vast amounts of taxpayer money. To win the hearts and minds
%of the people, historical reconstructions are used precisely because
%they can be very persuasive.
Consider Figure \ref{fig:ipcc} which was featured prominently in the
Intergovernmental Panel on Climate Change report [\citet{IPCC01}] in the
summary for policy makers.\footnote{Figure \ref{fig:ipcc} appeared in
\citet{IPCC01} and is due to \citet{MaBrHu99} which is in turn based on
the analysis of multiple proxies pioneered by \citet{MaBrHu98}. Figure
\ref{fig:thousand} is a ``spaghetti graph'' of multiple reconstructions
appearing in \citet{Mannetal08}. Figure \ref{fig:nrc} appeared in \citet{NRC06}.}
The sharp upward slope of the graph in the late 20th century
is visually striking, easy to comprehend, and likely to alarm. The IPCC
report goes even further:

\begin{quote}
Uncertainties increase in more distant times and are always
much larger than in the instrumental record due to the use of
relatively sparse
proxy data. Nevertheless the rate and duration of warming of the 20th
century has been much greater than in any of the previous nine
centuries. Similarly, \textbf{it is likely that the 1990s have been the
warmest decade and 1998 the warmest year of the millennium.} [Emphasis added]
\end{quote}

\noindent Quotations like the above and graphs like those in Figures \ref
{fig:ipcc}--\ref{fig:nrc} are featured
prominently not only in official documents like the IPCC report but
also in widely viewed television programs [\citet{BBC08}], in film [\citet{Gore06}],
and in museum expositions [\citet{Roth08}], alarming both the
populace and policy makers.

%f1 ###

%%\includegraphics[width=4in]{fig_1000comparisons.pdf}
%last 1,000 years according to various older articles (bluish lines),
%newer articles (reddish lines), and instrumental record (black line).
%Detailed references for each reconstruction can be found here:

%f2 ###

%f3 ###

It is not necessary to know very much about the underlying methods to
see that graphs such as Figure \ref{fig:ipcc} are problematic as
descriptive devices. First, the superposition of the instrumental
record (red) creates a strong but entirely misleading contrast. The
blue historical reconstruction is necessarily smoother with less
overall variation than the red instrumental record since the
reconstruction is, in a broad sense, a \textit{weighted average} of \textit{all}
global temperature histories conditional on the observed proxy
record. Second, the blue curve closely matches the red curve during the period
1902 AD to 1980 AD because this period has served as the training
data and therefore the blue curve is calibrated to the red during it
(note also the red curve is plotted from 1902 AD to 1998 AD). This sets up the
erroneous visual expectation that the reconstructions are more accurate than
they really are. A careful viewer would know to temper such expectations
by paying close attention to the reconstruction error bars given by the wide
gray regions. However, even these are misleading because these are, in fact,
\textit{pointwise} confidence intervals and not confidence curves for the
entire \textit{sample path} of surface temperature. Furthermore, the gray
regions themselves fail to account for model uncertainty.

%s2 ###
\section{Controversy}\label{controversy}

%There are two underlying statistical problems that must be confronted
%in order to reconstruct temperatures from proxy data: (i) the
%relationship between the proxies and temperature is (surprisingly)
%weak and (ii) in order to extract a stronger signal a large number of
%climate covariates are combined; there are more covariates than there
%are years in the instrumental (training) period. Both these issues are
%the source of controversy. Since the signal is so weak (for example,
%no proxy record, anywhere on earth, significantly tracks even the sign
%of annual temperature changes), the case must be built entirely on
%statistical methodologies whose complexity is assured by the brevity
%of the instrumental period and the number and size of the proxy record.

With so much at stake both financially and ecologically, it is not
surprising that these analyses have provoked several controversies.
While some have recently erupted in the popular press [\citet
{Jolis09}, \citet{John09}, \citet{JohnNaik09}], we root our discussion of
these controversies and their history as they unfolded in the academic and
scientific literature.

The first major controversy erupted when McIntyre and McKitrick (M\&M)
successfully replicated the \citet{MaBrHu98} study
[McIntyre and McKitrick (\citeyear{McIMcK03}, \citeyear{McIMcK05b}, \citeyear{McIMcK05})].
M\&M observed that the original \citet{MaBrHu98}
study (i) used only one principal component of the proxy record and
(ii) calculated the principal components in a ``skew''-centered fashion
such that they were centered by the mean of the proxy data over the
instrumental period (instead of the more standard technique of
centering by the mean of the entire data record). Given that the proxy
series is itself auto-correlated, this scaling has the effect of
producing a first principal component which is hockey-stick shaped
[\citet{McIMcK03}] and, thus, hockey-stick shaped temperature
reconstructions. That is, the very \textit{method} used in \citet{MaBrHu98}
guarantees the shape of Figure \ref{fig:ipcc}. M\&M made a further
contribution by applying the \citet{MaBrHu98} reconstruction methodology
to principal components computed in the standard fashion. The resulting
reconstruction showed a rise in temperature in the medieval period,
thus eliminating the hockey stick shape.

%Mann and his colleagues vigorously responded to M\&M with the
%observation that the hockey stick re-emerges when more principal
%components are retained \cite[]{MaBrHu04}. Furthermore, they argued,
%if one selects the number of retained principal components through
%cross-validation on \textit{two blocks} of heldout instrumental
%temperature records (i.e., the first 50 years of the instrumental
%period and the last 50 years), then enough principal components are
%retained such that the hockey stick re-emerges. Since the hockey stick
%is the shape selected by validation, climate scientists argue it is
%the correct one\footnote{Climate scientists call such reconstructions
%"more skilled." Statisticians would say they have lower out-of-sample
%root mean square error.}.

Mann and his colleagues vigorously responded to M\&M to justify the
hockey stick [\citet{MaBrHu04}]. They argued that one should not limit
oneself to a single principal component as in \citet{MaBrHu98}, but,
rather, one should select the number of principal components retained
through cross-validation on \textit{two blocks} of heldout instrumental
temperature records (i.e., the first 50 years of the instrumental
period and the last 50 years). When this procedure is followed, four
principal components are retained, and the hockey stick re-emerges even
when the PCs are calculated in the standard fashion. Since the hockey
stick is the shape selected by validation, climate scientists argue it
is therefore the correct one.\footnote{Climate scientists call such
reconstructions ``more skilled.'' Statisticians would say they have
lower out-of-sample root mean square error. We take up this subject in
detail in Section \ref{validation}.}

The furor reached such a level that Congress took up the matter in
2006.\eject The Chairman of the Committee on Energy and Commerce and that of
the~Subcommittee on Oversight and Investigations formed an \textit{ad
hoc} committee of statisticians to review the findings of M\&M. Their
Congressional report [\citet{WegScoSaid06}] confirmed M\&M's
finding regarding skew-centered principal components (this finding was
yet again confirmed by the National Research Council
[\citet{NRC06}]).

In his Congressional testimony [\citet{Wegman06}], committee
chairman Edward Wegman excoriated \citet{MaBrHu04} for use of additional
principal components beyond the first after it was shown that their
method led to spurious results:

\begin{quote}
\vspace{2pt}
In the MBH original, the hockey stick emerged in PC1 from the
bristlecone/\break foxtail pines. If one centers the data properly the hockey
stick does not emerge until PC4. Thus, a substantial change in strategy
is required in the MBH reconstruction in order to achieve the hockey
stick, a strategy which was specifically eschewed in MBH\ldots\ a cardinal
rule of statistical inference is that the method of analysis must be
decided before looking at the data. The rules and strategy of analysis
cannot be changed in order to obtain the desired result. Such a
strategy carries no statistical integrity and cannot be used as a basis
for drawing sound inferential conclusions.
\vspace{2pt}
\end{quote}

\noindent Michael Mann, in his rebuttal testimony before Congress, admitted to
having made some questionable choices in his early work. But, he
strongly asserted that none of these earlier problems are still
relevant because his original findings have been confirmed again and
again in subsequent peer reviewed literature by large numbers of highly
qualified climate scientists using vastly expanded data records [e.g.,
\citet{ManRut02}, \citet{Luteretal04}, \citeauthor{MaRuWaAm05} (\citeyear{MaRuWaAm05},
\citeyear{MaRuWaAm07}, \citeyear{Mannetal08}),
\citet{Ruthetal05}, \citet{WahAmm06}, \citet{WahRitAmm06}, \citet{LiNycAmm07}] even
if criticisms do exist [e.g., \citet{Vonetal04}].

%Very recently, the popular press has reported on the huge
%uncertainties associated with climate reconstructions and the
%continued controversy between M\&M and Mann \textit{et al.}
%scandal, which erupted in November 2009 when the email server of the
%Climatic Research Unit (CRU) of the University of East Anglia (UEA)
%was hacked and over 1,000 emails and more than 2,000 other documents
%were stolen (\citet{John09}, \citet{JohnNaik09}). The emails suggest
%collusion among researchers advocating AGW, contain discussions of how
%to rebut AGW skeptics and keep them out of peer-reviewed journals, and
%talk of deleting files to prevent data being revealed under the
%Freedom of Information Act.

The degree of controversy associated with this endeavor can perhaps be
better understood by recalling Wegman's assertion that there are very
few mainstream statisticians working on climate reconstructions [\citet{WegScoSaid06}].
This is particularly surprising not only because the
task is highly statistical but also because it is extremely difficult.
The data is spatially and temporally autocorrelated. It is massively
incomplete. It is not easily or accurately modeled by simple
autoregressive processes. The signal is very weak and the number of
covariates greatly outnumbers the number of independent observations of
instrumental temperature. Much of the analysis in this paper explores
some of the difficulties associated with model selection and prediction
in just such contexts. We are not interested at this stage in engaging
the issues of data quality. To wit, henceforth and for the remainder of
the paper, we work entirely with the data from \citet{Mannetal08}.\looseness=1\footnote{In the sequel,
we provide a link to \textit{The Annals of Applied Statistics} archive which hosts the data
and code we used for this paper. The \citet{Mannetal08} data can be found at
\url{http://www.meteo.psu.edu/\textasciitilde mann/supplements/MultiproxyMeans07/}.
However, we urge caution because this website is periodically updated
and therefore may not match the data we used even though at one time it
did. For the purposes of this paper, please follow our link to \textit{The
Annals of Applied Statistics} archive.}

This is by far the most comprehensive publicly available database of
temperatures and proxies collected to date. It contains 1209 climate
proxies (with some going back as far as 8855 BC and some continuing up
till 2003 AD). It also contains a database of eight global annual
temperature aggregates dating 1850--2006 AD (expressed as deviations or
``anomalies'' from the 1961--1990 AD average\footnote{For details, see
\url{http://www.cru.uea.ac.uk/cru/data/temperature/.}}). Finally,
there is a database of 1732 local annual temperatures dating
1850--2006 AD (also expressed as anomalies from the 1961--1990 AD
average).\footnote{The \citet{Mannetal08} original begins with the
HadCRUT3v local temperature data given in the previous link.
Temperatures are given on a five degree longitude by five degree
latitude grid. This would imply 2592 cells in the global grid. \citet
{Mannetal08} disqualified 860 such cells because they contained less
than 10\% of the annual data thus leaving 1732.} All three of these
datasets have been substantially processed including smoothing and
imputation of missing data [\citet{Mannetal08}]. While these present
interesting problems, they are not the focus of our inquiry. We \textit{assume}
that the data selection, collection, and processing performed
by climate scientists meets the standards of their discipline. Without
taking a position on these data quality issues, we thus take the
dataset as given. We further make the assumptions of linearity and
stationarity of the \textit{relationship} between temperature and proxies,
an assumption employed throughout the climate science literature [\citet{NRC06}]
noting that ``the stationarity of the relationship does not
require stationarity of the series themselves'' [\citet{NRC06}]. Even
with these substantial assumptions, the paleoclimatological
reconstructive endeavor is a very difficult one and we focus on the
substantive \textit{modeling} problems encountered in this
setting.\looseness=1

Our paper structure and major results are as follows. We first discuss
the strength of the proxy signal in this $p \gg n$ context (i.e., when
the number of covariates or parameters, $p$, is much larger than the
number of datapoints, $n$) by comparing the performance, in terms of
holdout RMSE, of the proxies against several alternatives. Such an
exercise is important because, when $p \gg n$, there is a sizeable risk
of overfitting and in-sample performance is often a poor benchmark for
out-of-sample performance. We will show that the proxy record easily
does better at predicting out-of-sample global temperature than simple
rapidly-mixing stationary processes generated independently of the true
temperature record. On the other hand, the proxies do not fare so well
when compared to predictions made by more complex processes also
generated independently of any climate signal. That is, randomly
generated sequences are as ``predictive'' of holdout temperatures as
the proxies.

Next, we show that various models for predicting temperature can
perform similarly in terms of cross-validated out-of-sample RMSE \textit{but} have
very different historical temperature backcasts. Some of
these backcasts look like hockey sticks while others do not. Thus,
cross-validation is inadequate on its own for model and backcast selection.

Finally, we construct and fit a full probability model for the
relationship between the 1000-year-old proxy database and Northern
Hemisphere average temperature, providing appropriate \textit{pathwise}
standard errors which account for parameter uncertainty. While our
model offers support to the conclusion that the 1990s were the warmest
decade of the last millennium, it does not predict temperature as well
as expected even in-sample. The model does much worse on contiguous
30-year holdout blocks. Thus, we remark in conclusion that natural
proxies are severely limited in their ability to predict average
temperatures and temperature gradients.

%All data and code used in this paper are available at the following
%website: \\
All data and code used in this paper are provided in the supplementary
materials [\citet{McSWyn11sup}].
%at \textit{The Annals of Applied Statistics} supplementary materials
%website: %\\

%s3 ###
\section{Model evaluation}\label{validation}

%s3.1 ###
\subsection{Introduction}

A critical difficulty for paleoclimatological reconstruction is that
the temperature signal in the proxy record is surprisingly weak. That
is, very few, if any, of the individual natural proxies, at least those
that are uncontaminated by the documentary record, are able to explain
an appreciable amount of the \textit{annual} variation in the local
instrumental temperature records. Nevertheless, the proxy record is
quite large, creating an additional challenge: there are many more proxies
than there are years in the instrumental temperature record. In this
setting, it is easy for a model to overfit the comparatively short
instrumental record and therefore model evaluation is especially
important. Thus, the main goals of this section are twofold. First, we
endeavor to judge regression-based methods for the specific task of
predicting blocks of temperatures in the instrumental period. Second,
we study specifically how the determination of statistical significance
varies under different specifications of the null distribution.
% BLAKE: SHOULD WE ADD MORE HERE???

Because the number of proxies is much greater than the number of years
for which we have temperature data, it is unavoidable that some type of
dimensionality reduction is necessary even if there is no principled
way to achieve this. As mentioned above, early studies
[\citeauthor{MaBrHu98} (\citeyear{MaBrHu98}, \citeyear{MaBrHu99})] used principal
components analysis for this purpose.
Alternatively, the number of proxies can be lowered through a threshold
screening process [\citet{Mannetal08}] whereby each proxy sequence is
correlated with its closest local temperature series and only those
proxies whose correlation exceeds a given threshold are retained for
model building. This is a reasonable approach, but, for it to offer
serious protection from overfitting the temperature sequence, it is
necessary to detect ``spurious
correlations.''

%f4 ###
\begin{figure}[b]

\includegraphics{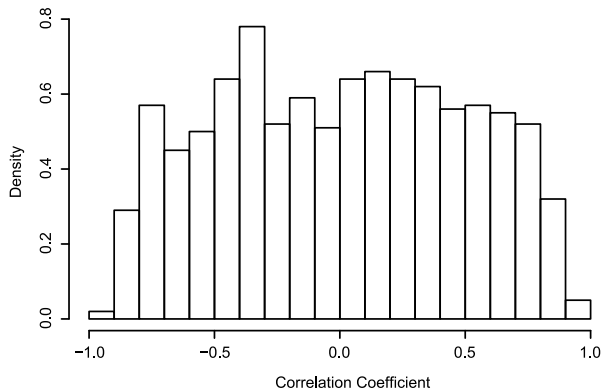}

\caption{Simulated sample correlation coefficient distribution of two
independent random walks. One thousand independent pairs of random
walks each of length $149$ were sampled to generate the above
histogram.}\label{fig:rwcorr}
\end{figure}

The problem of spurious correlation arises when one takes the
correlation of two series which are themselves highly autocorrelated
and is well studied in the time series and econometrics literature
[\citet{Yule26}, \citet{GraNew74}, \citet{Phillips86}]. When two independent time series
are nonstationary (e.g., random walk), locally nonstationary (e.g.,
regime switching), or strongly autocorrelated, then the distribution of
the empirical correlation coefficient is surprisingly variable and is
frequently large in absolute value (see Figure \ref{fig:rwcorr}).
Furthermore, standard model statistics (e.g., \textit{t}-statistics) are
inaccurate and can only be corrected when the underlying stochastic
processes are both known and modeled (and this can only be done for
special cases).
%A potential solution to the problem of spurious correlation is
%out-of-sample validation on a reserved holdout block of data.

%f5 ###
\begin{figure}

\includegraphics{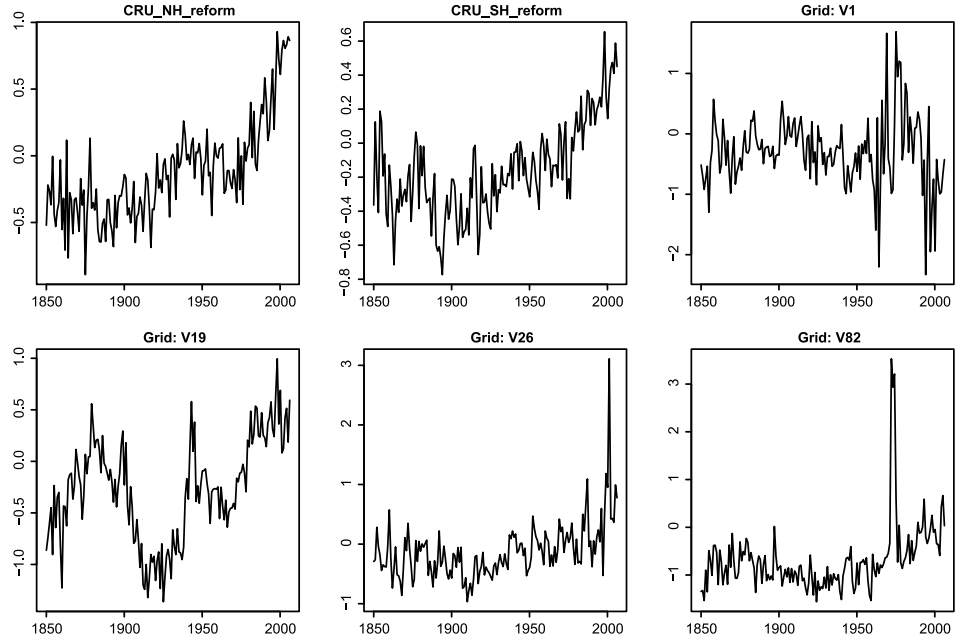}

\caption{CRU Northern Hemisphere annual mean land temperature, CRU
Southern Hemisphere annual mean land temperature, and four local
temperatures the grids of which contain \textup{(i)} Spitsbergen island in
the Svalbard archipelago in the Artic, \textup{(ii)} the north portion of
the Omsk oblast in southwestern Siberia, \textup{(iii)} Attu Island, the
westernmost island in the Aleutian islands arcipelago, and \textup{(iv)}
Baysuat in the Aktobe Province, Kazakhstan. The x-axis gives the year
and the y-axis gives the temperature anomaly from 1961--1990 AD
average in degrees Celsius.}\label{fig:temp}
\end{figure}

\begin{figure}

\includegraphics{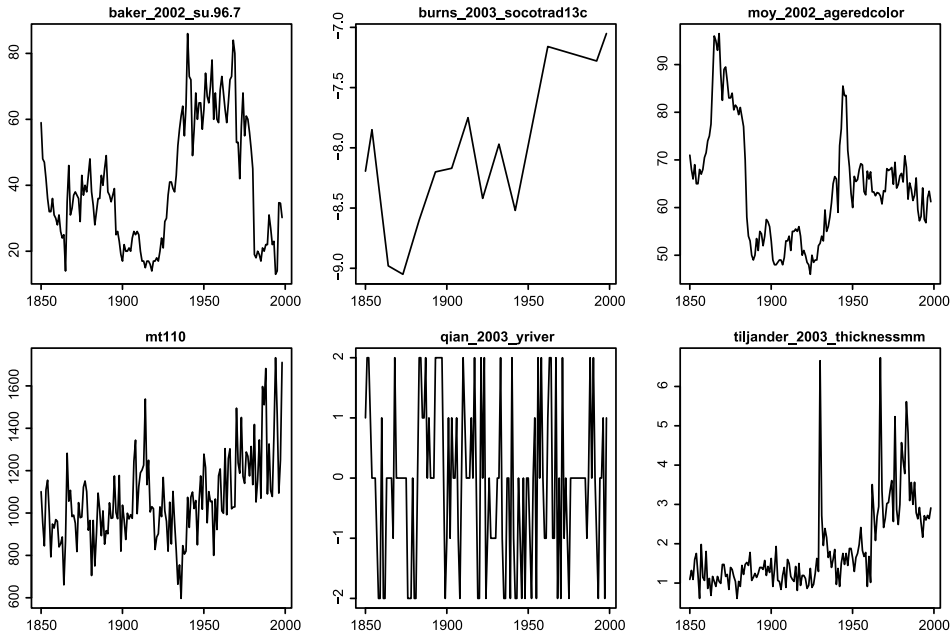}

\caption{Six proxy time series plotted during the instrumental period:
speleothems in Scotland, monsoons in India, lake sediment in Ecuador;
tree rings in Montana, dry/wet variation on the Yellow River, and lake
sediments in Finland.}\label{fig:proxy}\vspace*{-0.5pt}
\end{figure}

\begin{figure}

\includegraphics{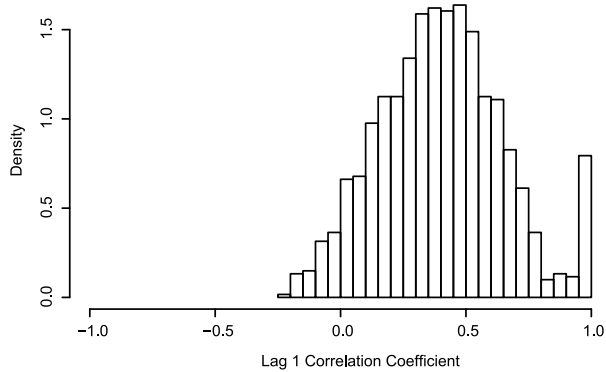}

\caption{Sample lag one autocorrelation coefficient for the 1209
proxies during the instrumental period.}\label{fig:proxycorr}
\end{figure}

As can be seen in Figures \ref{fig:temp} and \ref{fig:proxy}, both the
instrumental temperature record as well as many of the proxy sequences
are not appropriately modeled by low order stationary autoregressive
processes. The dependence structure in the data is clearly complex and
quite evident from the graphs. More quantitatively, we observe that the
sample first-order autocorrelation of the CRU Northern Hemisphere
annual mean land temperature series is nearly 0.6 (with significant \textit{partial}
autocorrelations out to lag four). Among the proxy sequences,
a full one-third have empirical lag one autocorrelations of at least 0.5
(see Figure \ref{fig:proxycorr}). Thus, standard correlation
coefficient test statistics are not reliable measures of significance
for screening proxies against local or global temperatures series. A
final more subtle and salient concern is that, if the screening process
involves the entire instrumental temperature record, it corrupts the
model validation process: no subsequence of the temperature series can
be truly considered out-of-sample.

To solve the problem of spurious correlation, climate scientists have
used the technique of out-of-sample validation on a reserved holdout
block of data. The performance of any given reconstruction can then be
benchmarked and compared to the performance of various null models.
This will be our approach as well. However, we extend their validation
exercises by (i) expanding the class of null models and (ii)
considering interpolated holdout blocks as well as extrapolated ones.

%Climate scientists have generally chosen to evaluate their models
%using the first and last blocks of temperatures in the instrumental
%record. For a given temperature reconstruction, an out-of-sample RMSE
%of temperatures is computed. The reconstruction is evaluated by
%benchmarking its skill against a class of competitors.

% Percent of proxy autocorrelations greater than:
%> mean(abs(rho) > .1) [1] 0.8990902
%> mean(abs(rho) > .2) [1] 0.7799835
%> mean(abs(rho) > .3) [1] 0.655914
%> mean(abs(rho) > .4) [1] 0.4954508
%> mean(abs(rho) > .5) [1] 0.3333333
%> mean(abs(rho) > .6) [1] 0.2026468
%> mean(abs(rho) > .7) [1] 0.1058726
%> mean(abs(rho) > .8) [1] 0.05707196
%> mean(abs(rho) > .9) [1] 0.04549214

%s3.2 ###
\subsection{Preliminary evaluation}\label{initialval}

In this subsection, we discuss our validation scheme and compare the
predictive performance of the proxies against two simple models which
use only temperature itself for forecasting, the in-sample mean and
ARMA models. We use as our response $y_t$ the CRU Northern Hemisphere
annual mean land temperature. $X=\{x_{tj}\}$ is a centered and scaled
matrix of 1138 of the 1209 proxies, excluding the 71 Lutannt series
found in \citet{Luteretal04}.\footnote{These Lutannt ``proxies'' are
actually reconstructions calibrated to local temperatures in Europe and
thus are not true natural proxies. The proxy database may contain other
nonnatural proxies though we do not believe it does. The qualitative
conclusions reached in this section hold up, however, even when all 1209
proxies are used.} We use the years 1850--1998 AD for these tests
because very few proxies are available after 1998 AD.\footnote{Only 103
of the 1209 proxies are available in 1999 AD, 90 in 2000 AD, eight in
2001 AD, five in 2002 AD, and three in 2003 AD.}

%f6 ###

To assess the strength of the relationship between the natural proxies
and temperature, we cross-validate the data. This is a standard
approach, but our situation is atypical since the temperature sequence
is highly autocorrelated. To mitigate this problem, we follow the
approach of climate scientists in our \textit{initial} approach and fit
the instrumental temperature record using \textit{only} proxy covariates.
Nevertheless, the errors and the proxies are temporally correlated
which implies that the usual method of selecting random holdout sets
will not provide an effective evaluation of our model. Climate
scientists have instead applied ``block'' validation, holding out two
contiguous blocks of instrumental temperatures: a ``front'' block
consisting of the first 50 years of the instrumental record and a
``back'' block consisting of the last 50 years.

%f7 ###

On the one hand, this approach makes sense since our ultimate task is
to extrapolate our data backward in time and only the first and last
blocks can be used for this purpose specifically. On the other hand,
limiting the validation exercise to these two blocks is problematic
because both blocks have very dramatic and obvious features: the
temperatures in the initial block are fairly constant and are the
coldest in the instrumental record, whereas the temperatures in the
final block are rapidly increasing and are the warmest in the
instrumental record. Thus, validation conducted on these two blocks
will \textit{prima facie} favor procedures which project the local level
and gradient of the temperature near the boundary of the in-sample
period. However, while such procedures perform well on the front and
back blocks, they are not as competitive on interior blocks.
Furthermore, they cannot be used for plausible historical
reconstructions! A final serious problem with validating on only the
front and back blocks is that the extreme characteristics of these
blocks are widely known; it can only be speculated as to what extent
the collection, scaling, and processing of the proxy data as well as
modeling choices have been affected by this knowledge.\vadjust{\goodbreak}
%BLAKE: MCINT has been SCREAMING about this. Do we mention it?

Our approach is to consecutively select all possible contiguous blocks
for holding out. For example, we take a given contiguous 30-year block
from the 149-year instrumental temperature record (e.g., 1900--1929 AD)
and hold it out. Using only the remaining 119 years (e.g, 1850--1899 AD
and 1930--1998 AD), we tune and fit our model. Finally, we then use
the fitted model to obtain predictions for each of the 30 years in the
holdout block and then calculate the RMSE on this block.

We then repeat the procedure outlined in the previous paragraph over
all 120 possible contiguous holdout blocks in order to approximate
the\break distribution of the holdout RMSE that is expected using this
procedure.\footnote{We performed two variations of this procedure. In
the first variation, we continued to hold out 30 years; however, we
calculated the RMSE for only the middle 20 years of the 30-year holdout
block, leaving out the first five and last five years of each block in
order to reduce the correlation between holdout blocks. In the second
variation, we repeated this procedure using 60-year holdout blocks. In
both cases, all qualitative conclusions remained the same. Considering
smaller holdout blocks such as 15 years could be an interesting
extension. However, over such short intervals, the global temperature
series itself provides substantial signal even without the use of
proxies. Furthermore, given the small size of the dataset and lack of
independence between 15-, 30-, and 60-year holdout blocks, this might
raise concerns about overfitting and over-interpreting the data.} We
note this test only gives a sense of the ability of the proxies to
predict the \textit{instrumental temperature record} and it says little
about the ability of the proxies to predict temperature several hundred
or thousand years back. Climate scientists have argued, however, that
this long-term extrapolation is scientifically legitimate [\citet{MaBrHu98}, \citet{NRC06}].

Throughout this section, we assess the strength of the proxy signal by
building models for temperature using the Lasso [\citet{Tib96}]. The
Lasso is a penalized least squares method which selects
\[
\hat\beta^{\mathrm{Lasso}} = \argmin\limits_{\beta} \Biggl\{ \sum_{i=1}^{n} \Biggl(y_i - \beta_0 -
\sum_{j=1}^{p} x_{ij} \beta_j\Biggr)^2 +
\lambda\sum_{i=1}^{p} |\beta_i| \Biggr\}.
\]
As can be seen, the intercept $\beta_0$ is not penalized. Typically
(and in this paper), the matrix of predictors $X$ is centered and
scaled, and $\lambda$ is chosen by cross-validation. Due to the $L_1$
penalty, the Lasso tends to choose sparse $\hat\beta^{\mathrm{Lasso}}$, thus
serving as a variable selection methodology and alleviating the $p \gg
n$ problem. Furthermore, since the Lasso tends to select only a few of
a set of correlated predictors, it also helps reduce the problem of
spatial correlation among the proxies.

We select the Lasso tuning parameter $\lambda$ by performing ten
repetitions of five-fold cross-validation on the 119 in-sample years
and choosing the value $\lambda= \hat\lambda$ which provides the best
RMSE. We then fit the Lasso to the full 119-year in-sample dataset
using $\lambda= \hat\lambda$ to obtain $\hat\beta^{\mathrm{Lasso}}$. Finally,
we can use $\hat\beta^{\mathrm{Lasso}}$ to obtain predictions for each of the
30 years in the holdout block and then calculate the RMSE on this block.

We chose the Lasso because it is a reasonable procedure that has proven
powerful, fast, and popular, and it performs comparably well in a $p
\gg n$ context. Thus, we believe it should provide predictions which
are as good or better than other methods that we have tried (evidence
for this is presented in Figure \ref{fig:alt2}). Furthermore, we are as
much interested in how the proxies fare as predictors when varying the
holdout block and null distribution (see Sections \ref{pseudoval} and
\ref{interpextrap}) as we are in performance. In fact, all analyses in
this section have been repeated using modeling procedures other than
the Lasso and qualitatively all results remain more or less the same.

%%\includegraphics[width=3.5in]{fig_proxy_mean.pdf}
%blocks: the RMSE for the model of the proxies chosen by the Lasso is
%plotted on the y-axis and the RMSE of the sample mean is plotted on
%the x-axis.}\label{fig:proxymean}

As an initial test, we compare the holdout RMSE using the proxies to
two simple models which only make use of temperature data, the
in-sample mean and ARMA models. First, the proxy model and the
in-sample mean seem to perform fairly similarly, with the proxy-based
model beating the sample mean on only 57\% of holdout blocks. A
possible reason the sample mean performs comparably well is that the
instrumental temperature record has a great deal of annual variation
which is apparently uncaptured by the proxy record. In such settings, a
biased low-variance predictor (such as the in-sample mean) can often
have a lower out-of-sample RMSE than a less biased but more variable
predictor. Finally, we observe that the performance on different
validation blocks are not independent, an issue which we return to
in Section \ref{interpextrap}.

We also compared the holdout RMSE of the proxies to another more
sophisticated model which, like the in-sample mean, only makes use of
temperature data and makes no reference to proxy data. For each holdout
block, we fit various ARMA($p$, $q$) models; we let $p$ and $q$ range
from zero to five and chose the values which give the best AIC. We then use
this model to forecast the temperature on the holdout block. This model
beats the proxy model 86\% of the time.

%f8 ###
\begin{figure}

\includegraphics{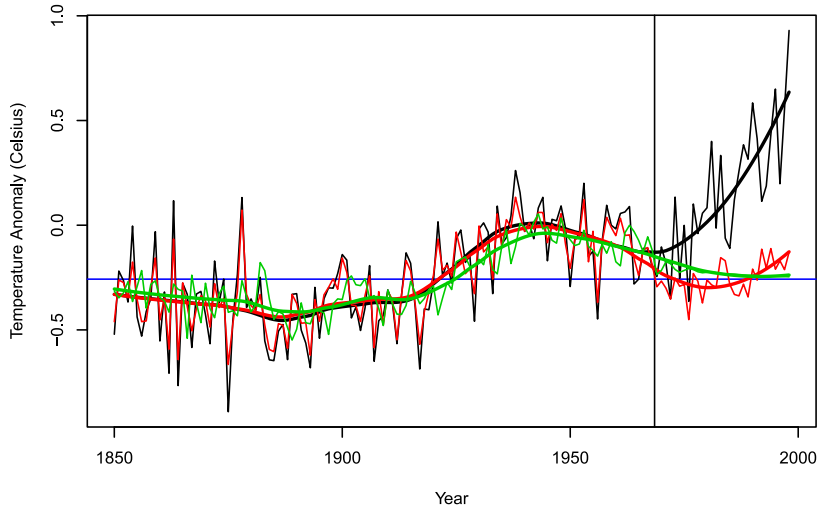}

\caption{CRU Northern Hemisphere annual mean land temperature is given
by the thin black line and a smoothed version is given by the thick
black line. The forecast produced by applying the Lasso to the proxies
is given by the thin red line and a smoothed version is given by the
thick red line. The in-sample mean is given by the horizontal blue
line. The forecast produced by ARMA modeling is given by the thin green
line and a smoothed version is given by the thick green line. The Lasso
and ARMA models and the mean are fit on 1850--1968 AD and
forecast on 1969--1998 AD.}\label{fig:last30}
\end{figure}

In Figure \ref{fig:last30}, we focus on one particular holdout block,
the last 30 years of the series.\footnote{In this and all subsequent
figures, smooths are created by using the \texttt{loess} function in R
with the span set to 0.33.} The in-sample mean and the ARMA model
completely miss the rising trend of the last 30 years; in fact, both
models are essentially useless for backcasting and forecasting since
their long-term prediction is equal to the in-sample mean. On the other
hand, the record of 1138 proxies does appear to capture the rising
trend in temperatures (in the sequel, we will assess the statistical
significance of this). Furthermore, the differences in temperature and
the differences in the proxy forecast are significantly correlated
($p=0.021$), with the same sign in 21 out of the 29 years ($p=0.026$).

%s3.3 ###
\subsection{Validation against pseudo-proxies}\label{pseudoval}

Because both the in-sample mean and the ARMA model always forecast the
mean in the long-term, they are not particularly useful models for the
scientific endeavor of temperature reconstruction. Furthermore, the
fact that the Lasso-selected linear combination of the proxies beats
the in-sample mean on 57\% of holdout blocks and the ARMA model on 14\%
of holdout blocks is difficult to interpret without solid benchmarks of
performance.

One way to provide benchmarks is to repeat the Lasso procedure outlined
above using 1138 ``pseudo-proxies'' in lieu of the 1138 real proxies.
That is, replace the natural proxies of temperature by an alternate set
of time series. Any function of the proxies, with their resultant
temperature reconstruction, can be validated by comparing the ability
of the proxies to predict out-of-sample instrumental temperatures to
the ability of the pseudo-proxies.

The use of pseudo-proxies is quite common in the climate science
literature where pseudo-proxies are often built by adding an AR1 time
series (``red noise'') to natural proxies, local temperatures, or
simulated temperatures generated from General Circulation Models [\citet{ManRut02}, \citet{WahAmm06}]. These pseudo-proxies determine whether a given
reconstruction is ``skillful'' (i.e., statistically significant). Skill
is demonstrated with respect to a class of pseudo-proxies when the true
proxies outperform the pseudo-proxies with high probability
(probabilities are approximated by simulation). In our study, we use an
even \textit{weaker} benchmark than those in the climate science
literature: our pseudo-proxies are random numbers generated \textit{completely independently} of the temperature series.

The simplest class of pseudo-proxies we consider are Gaussian White
Noise. That is, we apply the Lasso procedure outlined above to a $149
\times1138$ matrix of standard normal random variables. Formally, let
$\epsilon_t \iid N(0,1), t=1,2,\ldots.$ Then, our White Noise
pseudo-proxies are defined as $X_t \equiv\epsilon_t$ and we generate
1138 such series, each of length 149.

We also consider three classes of AR1 or ``red noise'' pseudo-proxies
since they are common in the climate literature [\citet{MaBrHu98},
\citet{Vonetal04}, \citet{Mannetal08}]. Again, if $\epsilon_t
\iid N(0,1)$, then an AR1 pseudo-proxy is defined as $X_t \equiv\phi
X_{t-1} + \epsilon_t$. Two of the classes are AR1 with the $\phi$
coefficient set in turn to 0.25 and 0.4 [these are the average
sample proxy autocorrelations reported in \citet{MaBrHu98} and
\citet{Mannetal08}, resp.]. The third class is more complicated.
First, we fit an AR1 model to each of the 1138 proxies and calculate
the sample AR1 coefficients\vspace*{1pt} $\hat\phi
_1,\ldots,\hat\phi_{1138}$. Then, we generate an AR1 series setting
$\phi= \hat\phi_i$ for each of these 1138 estimated coefficients. We
term this the empirical AR1 process. This approach is similar to that
of \citeauthor{McIMcK05b} (\citeyear{McIMcK05b}, \citeyear{McIMcK05c})
who use the full empirical autocorrelation function to generate
trend-less pseudo-proxies.

We also consider Brownian motion pseudo-proxies formed by taking the
cumulative sums of $N(0,1)$ random variables. That is, if $\epsilon_t
\iid N(0,1)$, then a Brownian motion pseudo-proxy is defined as $X_t
\equiv\sum_{j=1}^{t} \epsilon_j = X_{t-1} + \epsilon_t$.

White Noise and Brownian motion can be thought of as special cases of
AR1 pseudo-proxies. In fact, they are the \textit{extrema} of AR1
processes: White Noise is AR1 with the $\phi$ coefficient set to zero
and Brownian motion is AR1 with the $\phi$ coefficient set to one.

%The final pseudo-proxy process we consider is not a special case of
%AR1 but is a more complicated process that we term ``Pseudo-Bridge."
%Pseudo-Bridge, so named because it resembles Brownian Bridge, is
%generated by the following procedure. First, we generate ``knots" by
%cumulatively summing $Poisson(30)$ random variables. Then, at each
%knot, we generate ``fixed points" via $N(0,3^2)$ random variables (we
%also let one be a knot with a fixed point of zero). Finally, we
%connect the (knot, fixed point) pairs via Brownian Bridge. We consider
%this process because, unlike Brownian motion, this process is
%stationary and does not drift off.

%Formally, Pseudo-Bridge is the series $X_t$ created by the following
%procedure. Let $A_1=1$ and $A_k \iid Pois(30)$ for $k=1,2,...$. Define
%$B_k \equiv\sum_{j=1}^{k} A_j$. Now, there are two cases:
% \item If $t = B_k$ for some $k$, then draw $X_t \sim N(0,3^2)$ for
%$t>1$ (for simplicity, we set $X_1 \equiv0$).
% \item If $t \neq B_k$ for any $k$, then $B_j < t < B_k$ for some $j$
%where $k = j + 1$. In this case, draw $X_t \sim N(\mu_t,\sigma_t^2)$
%where $\mu_t = X_{B_j} + \frac{t - B_j}{B_k - B_j} (X_{B_k} -
%X_{B_j})$ and $\sigma_t^2 = \frac{(t - B_j)(B_k - t)}{B_k-B_j}$.

Before discussing the results of these simulations, it is worth
emphasizing why this exercise is necessary. That is, why can't one
evaluate the model using standard regression diagnostics (e.g., \textit{F}-statistics, \textit{t}-statistics, etc.)? One cannot because of two
problems mentioned above: (i) the $p \gg n$ problem and (ii) the fact
that proxy and temperature autocorrelation causes spurious correlation
and therefore invalid model statistics (e.g., \textit{t}-statistics). The
first problem has to be dealt with via dimensionality reduction; the
second can only be solved when the underlying processes are known (and
then only in special cases).

Given that we do not know the true underlying dynamics, the
nonparametric, prediction-based approach used here is valuable. We
provide a variety of benchmark pseudo-proxy series and obtain holdout
RMSE distributions. Since these pseudo-proxies are generated
independently of the temperature series, we know they cannot be truly
predictive of it. Hence, the real proxies---if they contain linear
signal on temperatures---should outperform our pseudo-proxies, at least
with high probability.

For any given class of pseudo-proxy, we can estimate the probability
that a randomly generated pseudo-proxy sequence outperforms the true
proxy record for predicting temperatures in a given holdout block. A
major focus of our investigation is the sensitivity of this
outperformance ``\textit{p}-value'' to various factors. We proceed in two
directions. We first consider the level and variability in holdout RMSE
for our various classes of pseudo-proxies marginally over all 120
holdout blocks. Second, since these 120 holdout blocks are highly
correlated with one another, we study how the holdout RMSE varies from
block to block in Section~\ref{interpextrap}.

%f9 ###
\begin{figure}

\includegraphics{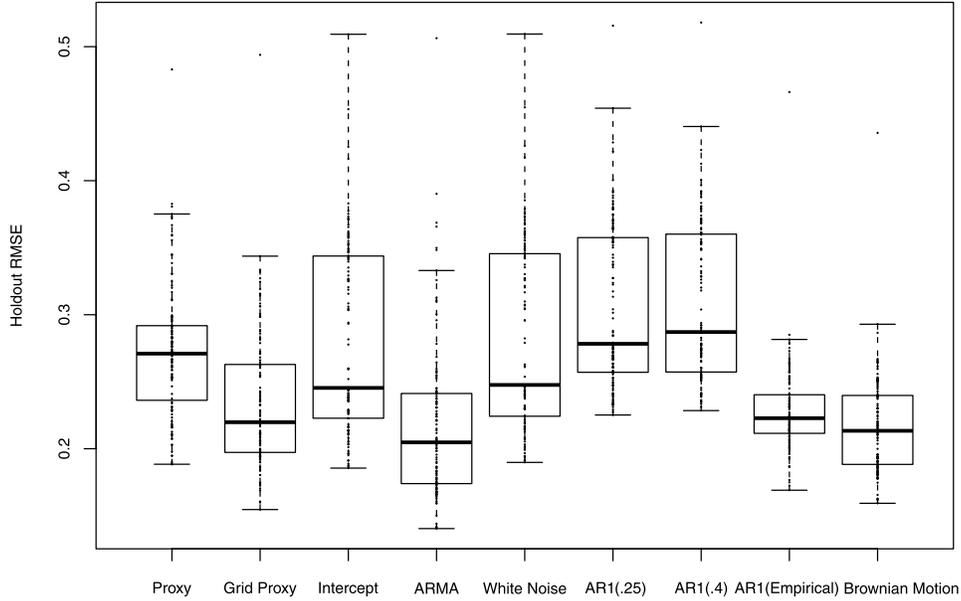}

\caption{Cross-validated RMSE on $30$-year holdout
blocks for various models fit to proxies and pseudo-proxies. The
procedures used to generate the Proxy, Intercept, and ARMA boxplots are
discussed in Section \protect\ref{initialval}. The procedures used to generate
the White Noise, AR1, and Brownian motion boxplots are discussed in
Section \protect\ref{pseudoval}. The procedure used to generate the Grid Proxy
boxplot is discussed in Section \protect\ref{grid}.}\label{fig:boxplot}
\end{figure}

We present our results in Figure \ref{fig:boxplot}, with the RMSE
boxplot for the proxies given first. As can be seen, the proxies have a
slightly worse median RMSE than the intercept-only model (i.e., the
in-sample mean) but the distribution is narrower. On the other hand,
the ARMA model is superior to both. When the Lasso is used on White
Noise pseudo-proxies, the performance is similar to the intercept-only
model because the Lasso is choosing a very parsimonious model.%, in fact,
%it frequently selects the intercept-only model.

The proxies seem to outperform the AR1(0.25) and AR1(0.4) models, with
both a better median and a lower variance. While this is encouraging,
it is also raises a concern: AR1(0.25) and AR1(0.4) are the models
frequently used as ``null benchmarks'' in the climate science
literature and they seem to perform worse than both the intercept-only
and White Noise benchmarks. This suggests that climate scientists are
using a particularly weak null benchmark to test their models. That the
null models may be too weak and the associated standard errors in
papers such as \citet{MaBrHu98} are not wide enough has already been
pointed out in the climate literature [\citet{Vonetal04}].\vadjust{\goodbreak} While there
was some controversy surrounding the result of this paper [\citet{WahRitAmm06}], its conclusions have been corroborated
[\citet{VonZor05}, \citet{Vonetal06}, \citet{LeZwTs08}, \citet{ChScTh09}].

Finally, the empirical AR1 process and Brownian motion both
substantially outperform the proxies. They each have a lower average holdout
RMSE and lower variability than that achieved by the proxies. This is
extremely important since these two classes of time series are
generated \textit{completely independently} of the temperature data. They
have \textit{no} long term predictive ability, and they cannot be used to
reconstruct historical temperatures. Yet, they significantly outperform
the proxies at 30-year holdout prediction!

In other words, our model performs better when using highly
autocorrelated noise rather than proxies to ``predict'' temperature.
The real proxies are less predictive than our ``fake'' data. While the
Lasso-generated reconstructions using the proxies are highly
statistically significant compared to simple null models, they do not
achieve statistical significance against sophisticated null models.

We are not the first to observe this effect. It was shown, in McIntyre and McKitrick (\citeyear{McIMcK05b}, \citeyear{McIMcK05c}),
that random sequences with complex local
dependence structures can predict temperatures. Their approach has been
roundly dismissed in the climate science literature:

\begin{quote}
To generate ``random'' noise series, MM05c apply the full
autoregressive structure of the real world proxy series. In this way,
they in fact train their stochastic engine with significant (if not
dominant) low frequency \textit{climate signal} rather than purely
nonclimatic noise and its persistence. [Emphasis in original]

\hfill\citet{AmmWah07}
\end{quote}

Broadly, there are two components to any climate signal. The first
component is the local time dependence made manifest by the strong
autocorrelation structure observed in the temperature series itself. It
is easily observed that short term future temperatures can be predicted
by estimates of the local mean and its first derivatives [\citet{GreArmSoo09}]. Hence, a procedure that fits sequences with complex
local dependencies to the instrumental temperature record will recover
the ability of the temperature record to self-predict in the short run.

The second component---long-term changes in the temperature
series---can, on the other hand, only be predicted by meaningful
covariates. The autocorrelation structure of the temperature series
does not allow for self-prediction in the long run. Thus,
pseudo-proxies like ours, which inherit their ability at short-term
prediction by borrowing the dependence structure of the instrumental
temperature series, have no more power to reconstruct temperature than
the instrumental record itself (which is entirely sensible since these
pseudo-proxies are generated independently of the temperature series).

\citet{AmmWah07} claim that significance thresholds set by Monte Carlo
simulations that use pseudo-proxies containing ``short term climate
signal'' (i.e., complex time dependence structures) are invalid:

\begin{quote}
Such thresholds thus enhance the danger of committing Type II errors
(inappropriate failure to reject a null hypothesis of no climatic
information for a reconstruction).
\end{quote}

\noindent We agree that these thresholds decrease power. Still, these thresholds
are the correct way to preserve the significance level. The proxy
record has to be evaluated in terms of its innate ability to
reconstruct historical temperatures (i.e., as opposed to its ability to
``mimic'' the local time dependence structure of the temperature
series). \citet{AmmWah07} wrongly attribute reconstructive skill to the
proxy record which is in fact attributable to the temperature record
itself. Thus, climate scientists are overoptimistic: the 149-year
instrumental record has significant local time dependence and therefore
far fewer independent degrees of freedom.

%This simulation would be the appropriate ``null benchmark" for the
%proxies if we knew they all followed an AR1 process (which is unlikely
%given the plots in Figure \ref{fig:proxy}).

%s3.4 ###
\subsection{Interpolation versus extrapolation}\label{interpextrap}

\begin{figure}

\includegraphics{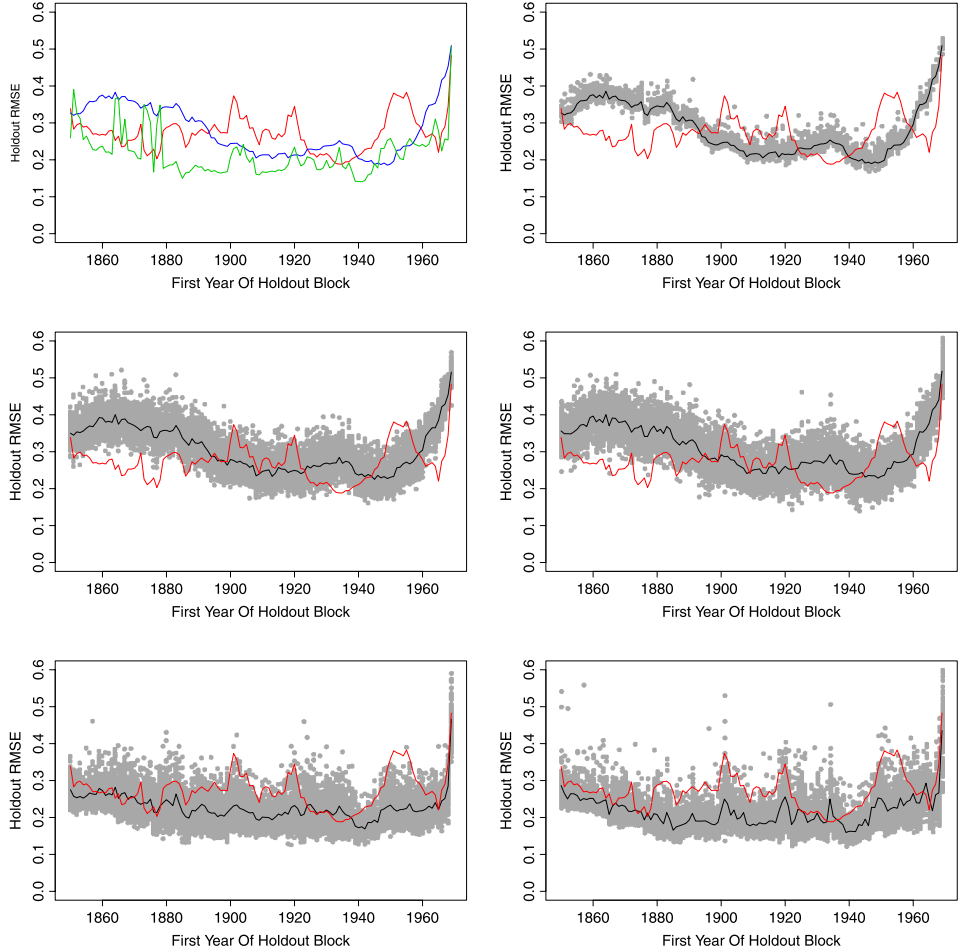}

\caption{Holdout RMSE by first year of holdout block. In all panels,
the Lasso-selected linear combination of the proxies is given in red.
In the upper-left panel, the in-sample mean is given in blue and the
ARMA model in green. In the upper-right panel, the average for the
White Noise pseudo-proxy is given in black. In the middle-left panel,
the average for the $\operatorname{AR1}(0.25)$ pseudo-proxy is given in black. In the
middle-right panel, the average for the $\operatorname{AR1}(0.4)$ pseudo-proxy is given
in black. In the lower-left panel, the average for the Empirical $AR1$
pseudo-proxy is given in black. In the lower-right panel, the average
for the Brownian motion pseudo-proxy is given in black. Confidence
intervals for the pseudo-proxies are given in gray and are formed by
taking $100$ samples of the pseudo-proxy matrix for each holdout
block.}\label{fig:rmseyr}
\vspace{-10pt}
\end{figure}

In our analysis, we expanded our set of holdout blocks to include all
contiguous 30-year blocks. The benefits of this are twofold. First,
this expansion increases our sample size from two (the front and back
blocks) to 120 (because there are 118 possible interior blocks).
Second, by expanding the set of holdout blocks, we mitigate the
potential effects of data snooping since salient characteristics of the
first and last blocks are widely known. On the other hand, this
expansion imposes difficulties. The RMSEs of overlapping blocks are
highly dependent. Furthermore, since temperatures are autocorrelated,
the RMSEs of neighboring nonoverlapping blocks are also dependent.
Thus, there is little new information in each block.\footnote{As noted
in a previous footnote, we considered a variation of our procedure
where we maintained 30-year holdout blocks but only calculated the RMSE
on the middle 20 years of the block, thus reducing the dependence
between overlapping and nearby blocks. All qualitative conclusions
remained the same.} We explore this graphically by plotting the RMSE of
each holdout block against the first year of the block in Figure~\ref
{fig:rmseyr}.

%f10 ###

We begin our discussion by comparing RMSE of the Lasso model fitted to
the proxies to RMSE of the in-sample mean and the RMSE of the ARMA
model in upper left panel of Figure \ref{fig:rmseyr}. As can be seen,
the ARMA model either dominates or is competitive on every holdout
block. The proxies, on the other hand, can match the performance of the
ARMA model only on the first 20 or so holdout blocks, but on other
blocks, they perform quite a bit worse.

More interesting is the examination of the performance of the
pseudo-proxies, as shown in the remaining five panels of Figure \ref
{fig:rmseyr}. In these graphs, we compare the RMSE of the proxies on
each holdout block to the RMSE of the pseudo-proxies. We also provide
confidence intervals for the pseudo-proxies at each block by simulating
100 draws of the pseudo-proxy matrix and repeating our fitting
procedure to each draw. As can be seen in the upper-right panel, the
proxies show statistically significant improvement over White Noise for
many of the early holdout blocks as well as many of the later ones.
However, there are blocks, particularly in the middle, where they
perform significantly worse.

When the AR1(0.25) and AR1(0.4) pseudo-proxies preferred by climate
scientists are used, the average RMSE on each is comparable to that
given by White Noise but the variation is considerably higher as shown
by the middle two panels of Figure \ref{fig:rmseyr}. Hence, the proxies
perform statistically significantly better on very few holdout blocks,
particularly those near the beginning of the series and those near the
end. This is a curious fact because the ``front'' holdout block and the
``back'' holdout block are the only two which climate scientists use to
validate their models. Insofar as this front and back performance is
anomalous, they may be overconfident in their results.

Finally, we consider the AR1 Empirical and Brownian motion
pseudo-proxies in the lower two panels of Figure \ref{fig:rmseyr}. For
almost all holdout blocks, these pseudo-proxies have an average RMSE
that is as low or lower than that of the proxies. Further, for no block
is the performance of true proxies statistically significantly better
than that of either of these pseudo-proxies. Hence, we cannot reject
the null hypothesis that the true proxies ``predict'' equivalently to
highly correlated and/or nonstationary sequences of random noise that
are independent of temperature.

A little reflection is in order. By cross-validating on interior
blocks, we are able to greatly expand the validation test set. However,
reconstructing interior blocks is an interpolation of the training
sequence and paleoclimatological reconstruction requires extrapolation
as opposed to interpolation. Pseudo-proxy reconstructions can only
extrapolate a climate trend accurately for a very short period and then
only insofar as the local dependence structure in the pseudo-proxies
matches the local dependence structure in the temperature series. That
is, forecasts from randomly generated series can extrapolate
successfully only by chance and for very short periods.

On the other hand, Brownian motions and other pseudo-proxies with
strong local dependencies are quite suited to interpolation since their
in-sample forecasts are fitted to approximately match the the training
sequence datapoints that are adjacent to the initial and final points
of a test block. Nevertheless, true proxies also have strong local
dependence structure since they are temperature surrogates and
therefore should similarly match these datapoints of the training
sequence. Furthermore, unlike pseudo-proxies, true proxies are \textit{not} independent of temperature (in fact, the scientific presumption is
that they are \textit{predictive} of it). Therefore, proxy interpolations
on interior holdout blocks should be expected to outperform
pseudo-proxy forecasts notwithstanding the above.

%s3.5 ###
\subsection{Variable selection: True proxies versus pseudo-proxies}

While the use of noise variables such as the pseudo-proxies is not
unknown in statistics, such variables have typically been used to
augment a matrix of covariates rather than to replace it. For example,
\citet{WuBoSt07} augment a matrix of covariates with noise variables in
order to tune variable selection methodologies. Though that is not our
focus, we make use of a similar approach in order to assess the the
degree of signal in the proxies.

We first augment the in-sample matrix of proxies with a matrix of
pseudo-proxies of the same size (i.e., replacing the $119\times1138$
proxy matrix with a matrix of size $119\times2276$ which consists of
the original proxies plus\vadjust{\goodbreak} pseudo-proxies). Then, we repeat the Lasso
cross-validation described in Section \ref{initialval}, calculate the
percent of variables selected by the Lasso which are pseudo-proxies,
and average over all 120 possible blocks. If the signal in the proxies
dominates that in the pseudo-proxies, then this percent should be
relatively close to zero.

%t1 ###
\begin{table}
\tablewidth=174pt
\caption{Percent of pseudo-proxies selected by the Lasso}\label{badpct}
\begin{tabular*}{\tablewidth}{@{\extracolsep{\fill}}lc@{}}
\hline
\textbf{Pseudo-proxy} & \textbf{Percent selected} \\
\hline
White Noise & 37.8\% \\
AR1(0.25) & 43.5\% \\
AR1(0.4) & 47.9\% \\
Empirical AR1 & 53.0\% \\
Brownian Motion & 27.9\% \\
%Pseudo-Bridge & 62.7\% \\
\hline
\end{tabular*}
\end{table}

Table \ref{badpct} shows this is far from the case. In general, the
pseudo-proxies are selected about as often as the true proxies. That
is, the Lasso does not find that the true proxies have substantially
more signal than the pseudo-proxies.

%s3.6 ###
\subsection{Proxies and local temperatures}\label{grid}

We performed an additional test which accounts for the fact that
proxies are local in nature (e.g., tree rings in Montana) and therefore
might be better predictors of local temperatures than global
temperatures. Climate scientists generally accept the notion of
``teleconnection'' (i.e., that proxies local to one place can be
predictive of climate in another possibly distant place). Hence, we do
not use a distance restriction in this test. Rather, we perform the
following procedure.

Again, let $y_t$ be the CRU Northern Hemisphere annual mean land
temperature where $t$ indexes each year from 1850--1998 AD, and let
$X=\{x_{tj}\}$ be the centered and scaled matrix of 1138 proxies from
1850--1998 AD where $t$ indexes the year and $j$ indexes each proxy.
Further, let $Z=\{z_{tj}\}$ to be the matrix of the 1732 centered and
scaled local annual temperatures from 1850--1998 AD where again $t$
indexes the year and $j$ indexes each local temperature.

As before, we take a 30-year contiguous block and reserve it as a
holdout sample. Our procedure has two steps:
\begin{enumerate}
\item Using the 119 in-sample years, we perform ten repetitions of
five-fold cross-validation as described in Section \ref{initialval}. In
this case, however, instead of using the proxies $X$ to predict $y$, we
use the local temperatures $Z$. As before, this procedure gives us an
optimal value for the tuning parameter $\hat\lambda$ which we can use
on all 119 observations of $y$ and $Z$ to obtain $\hat\beta^{\mathrm{Lasso}}$.
\item Now, for each $j$ such that $\hat\beta^{\mathrm{Lasso}}_{j} \neq0$, we
create a Lasso model for $z_{\cdot j}$. That is, we perform ten repetitions
of five-fold cross-validation as in Section \ref{initialval} but using
$X$ to predict $z_{\cdot j}$. Again, this procedure gives us an optimal
value for the tuning parameter $\hat\lambda_j$ which we can use on all
119 observations of $z_{\cdot j}$ and $X$ to obtain $\hat\beta^{\mathrm{Lasso},(j)}$.
\end{enumerate}

Similarly, we can predict on the holdout block using a two-stage
procedure. For each $j$ such that $\hat\beta^{\mathrm{Lasso}}_{j} \neq0$, we
apply $\hat\beta^{\mathrm{Lasso},(j)}$ to $X$\vspace*{-3pt} to obtain $\hat z_{\cdot j}$ in the 30
holdout years. Then, we apply $\hat\beta^{\mathrm{Lasso}}$ to the
collection of $\hat z_{\cdot j}$ in order to obtain $\hat y_t$ in the 30
holdout years. Finally, we calculate the RMSE on the holdout block and
repeat this procedure over all 120 possible holdout blocks.

As in Section \ref{initialval}, this procedure uses the Lasso to
mitigate the $p \gg n$ problem. Furthermore, since the Lasso is
unlikely to select correlated predictors, it also attenuates the
problem of spatial correlation among the local temperatures and
proxies. But, this procedure has the advantage of relating proxies to
local temperatures, a feature which could be advantageous if these
relationships are more conspicuous and enduring than those between
proxies and the CRU global average temperature. The same is also potentially true
\textit{mutatis mutandis} of the relationship between the local
temperatures and CRU.

The results of this test are given by the second boxplot in Figure \ref
{fig:boxplot}. As can be seen, this method seems to perform somewhat
better than the pure global method. However, it does not beat the
empirical AR1 process or Brownian motion. That is, random series that
are independent of global temperature are as effective or more
effective than the proxies at predicting global annual temperatures in
the instrumental period. Again, the proxies are not statistically
significant when compared to sophisticated null models.

%s3.7 ###
\subsection{Discussion of model evaluation}

We can think of four possible explanatory factors for what we have
observed. First, it is possible that the proxies are in fact too weakly
connected to global annual temperature to offer a substantially
predictive (as well as reconstructive) model over the majority of the
instrumental period. This is not to suggest that proxies are unable to
detect large variations in global temperature (such as those
that distinguish our current climate from an ice age).
Rather, we suggest it is possible that natural proxies cannot reliably
detect the small and largely unpredictable changes in annual
temperature that have been observed over the majority of the
instrumental period. In contrast, we have previously shown that the
proxy record has some ability to predict the final 30-year block, where
temperatures have increased most significantly, better than chance
would suggest.

A second explanation is that the Lasso might be a poor procedure to
apply to these data. This seems implausible both because the Lasso has
been used successfully in a variety of $p \gg n$ contexts and because
we repeated the analyses in this section using modeling strategies
other than the Lasso and obtained the same general results. On the
other hand, climate scientists have basically used three different
statistical approaches: (i) scaling and averaging (so-called
``Composite Plus Scale'' or CPS) [\citet{NRC06}], (ii) principal
component regression [\citet{NRC06}], and (iii) ``Errors in Variables''
(EIV) regression [\citet{Schneider01}, \citet{MaRuWaAm07}]. The EIV approach is
considered the most reliable and powerful. The approach treats
forecasting (or reconstruction) from a missing data perspective using
the Expectation--Maximization algorithm to ``fill-in'' blocks of missing
values. The EM core utilizes an EIV generalized linear regression which
addresses the $p \gg n$ problem using regularization in the form of a
ridge regression-like total sum of squares constraint (this is called
``RegEM'' in the climate literature [\citet{MaRuWaAm07}]). All of these
approaches are intrinsically linear, like Lasso regression, although
the iterative RegEM can produce nonlinear functions of the covariates.
Fundamentally, there are only theoretical performance guarantees for
i.i.d. observations, while our data is clearly correlated across time.
The EM algorithm in particular lacks a substantive literature on
accuracy and performance without specific assumptions on the nature of
missing data. Thus, it not obvious why the Lasso regression should be
substantively worse than these methods. Nevertheless, in subsequent
sections we will study a variety of different and improved model
variations to confirm this.

A third explanation is that our class of competitive predictors (i.e.,
the pseudo-proxies) may very well provide unjustifiably difficult
benchmarks as claimed by \citet{AmmWah07} and discussed in Section \ref
{pseudoval}. Climate scientists have calibrated their performance using
either (i) weak AR1 processes of the kind demonstrated above as
pseudo-proxies or (ii) by adding weak AR1 processes to local
temperatures, other proxies, or the output from global climate
simulation models. In fact, we have shown that the the proxy record
outperforms the former. On the other hand, weak AR1 processes
underperform even white noise! Furthermore, it is hard to argue that a
procedure is truly skillful if it cannot consistently outperform noise,
no matter how artfully structured. In fact, Figure \ref{fig:proxy}
reveals that the proxy series contain very complicated and highly
autocorrelated time series structures which indicates that our complex
pseudo-proxy competitors are not entirely unreasonable.

Finally, perhaps the proxy signal can be enhanced by smoothing various
time series before modeling. Smoothing seems to be a standard approach
for the analysis of climate series and is accompanied by a large body
of literature [\citeauthor{Mann04} (\citeyear{Mann04}, \citeyear{Mann08})]. Still, from a statistical
perspective, smoothing time series raises additional questions and
problems. At the most basic level, one has to figure out which series
should be smoothed: temperatures, proxies, or both. Or, perhaps, only
the forecasts should be smoothed in order to reduce the forecast
variance. A further problem with smoothing procedures is that there are
many methods and associated tuning parameters and there are no clear
data-independent and hypothesis-independent methods of selecting among
the various options. The instrumental temperature record is also very
well known so there is no way to do this in a ``blind'' fashion.
Furthermore, smoothing data exacerbates all of the statistical
significance issues already present due to autocorrelation: two
smoothed series will exhibit artificially high correlations and both
standard errors and \textit{p}-values require corrections (which are again
only known under certain restrictive conditions).

%s4 ###
\section{Testing other predictive methods}

%s4.1 ###
\subsection{Cross-validated RMSE}\label{altmodels}

In this section, we pursue alternative procedures, including regression
approaches more directly\vadjust{\goodbreak} similar to techniques used by climate
scientists. We shall see, working with a similar dataset, that various
fitting methods can have both (i) very similar contiguous 30-year
cross-validated instrumental period RMSE distributions \textit{and} (ii)
very different historical backcasts.

Again, we use as our response the CRU Northern Hemisphere annual mean
land temperature from 1850--1998 AD and augment it with the 1732 local
temperature series when required. However, since we are ultimately
interested in large-scale reconstructions, we limit ourselves in this
section to only those 93 proxies for which we have data going back over
1000 years.\footnote{There are technically 95 proxies dating back this
far but three of them (tiljander\_2003\_darksum, tiljander\_2003\_lightsum, and tiljander\_2003\_thicknessmm) are highly correlated with
one another. Hence, we omit the latter two. Again, qualitatively,
results hold up whether one uses the reduced set of 93 or the full set
of 95 proxies. However, using the full set can cause numerical
instability issues.} Hence, our in-sample dataset consists of the CRU
global aggregate, the 1732 local temperatures, and the 93 proxies from
1850--1998 AD and we apply the cross-validation procedure discussed in
Section \ref{initialval} to it. We can then examine backcasts on the
998--1849 AD period for which only the proxies are available. We expect
that our prediction accuracy during the instrumental period will decay
somewhat since our set of proxies is so much smaller. However, the
problem of millennial reconstructions is much more interesting both
statistically and scientifically. It is well known and generally agreed
that the several hundred years before the industrial revolution were a
comparatively cool ``Little Ice Age'' [\citet{Matthes39}, \citet{Lamb90}]. What
happened in the early Medieval period is much more controversial and
uncertain [\citet{Ladurie71}, \citet{IPCC01}].

%f11 ###
\begin{figure}

\includegraphics{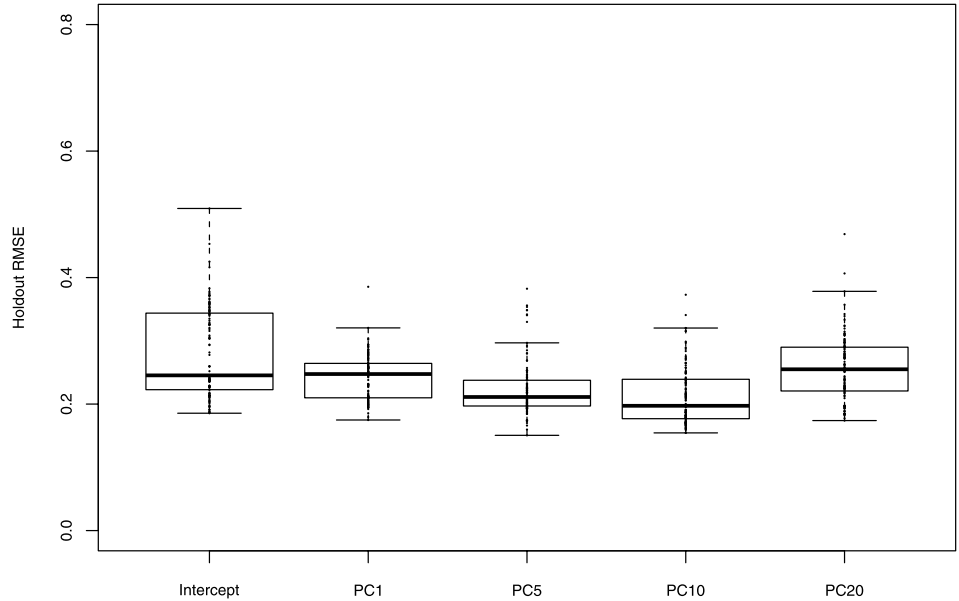}

\caption{Cross-validated RMSE on $30$-year holdout blocks for various
model specifications: intercept only and regression on the first one,
five, ten, and $20$ principal components of the proxies.}\label{fig:alt1}
\end{figure}

%We now examine how well the proxies predict under alternative model
%specifications. In the first set of studies, we examine RMSE
%distributions using an intercept-only model and OLS PC Regression on
%the first one, five, ten, and 20 principal components calculated from
%the full 1,001 x 93 proxy matrix. We also see how well OLS performs
%using all principal components. Finally, we examine how well the Lasso
%does on the ninety-three proxies as well as on the ninety-three
%principal components of the proxies.
% {\color{red} There is also BLASSO which chooses its tuning parameter
%via block-wise cross-validation...I think we should eliminate this
%technicality since it does not really matter...see plots}.
%Our results are shown in Figure \ref{fig:alt1}. As can be seen, all
%methods with the exception of OLS on the full set of principal
%components perform fairly similarly. The model using the first ten
%principal components as well as the Lasso model on the raw proxies
%seem to perform the best, but the results are fairly comparable.

We now examine how well the proxies predict under alternative model
specifications. In the first set of studies, we examine RMSE
distributions using an intercept-only model and ordinary least squares
regression on the first one, five, ten, and 20 principal components
calculated from the full $1001 \times93$ proxy matrix. Our results
are shown in Figure \ref{fig:alt1}. As can be seen, all of these
methods perform comparably, with five and ten principal component models
perhaps performing slightly better than the others.

%f12 ###
\begin{figure}

\includegraphics{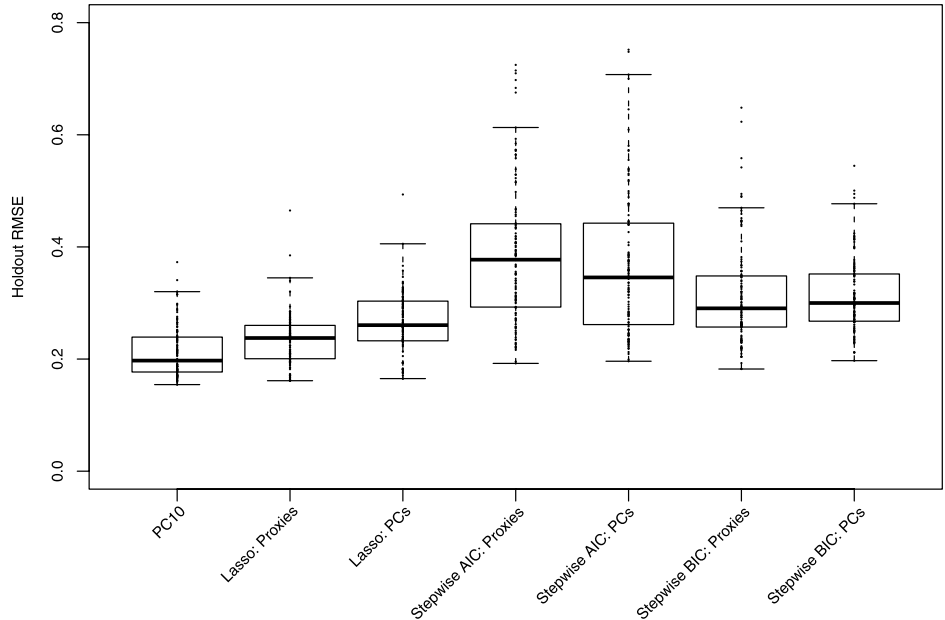}

\caption{Cross-validated RMSE on $30$-year holdout blocks for various
model specifications: regression on the first ten principal components
of the proxies, the Lasso applied to the proxies and the principal
components of the proxies, stepwise regression to maximize AIC applied
to the proxies and the principal components of the proxies, and
stepwise regression to maximize BIC applied to the proxies and the
principal components of the proxies.}\label{fig:alt2}
\end{figure}

In a second set of validations, we consider various variable selection
methodologies and apply them to both the raw proxies and the principal
components of the proxies. The methods considered are the Lasso and
stepwise regression designed to optimize AIC and BIC, respectively. We
plot our results in Figure \ref{fig:alt2} and include the boxplot of the ten principal component model
 from Figure \ref
{fig:alt1} for easy reference. As can be seen, the stepwise models
perform fairly similarly with one another. The Lasso performs slightly
better and predicts about as well as the ten principal component model.

%f13 ###
\begin{figure}

\includegraphics{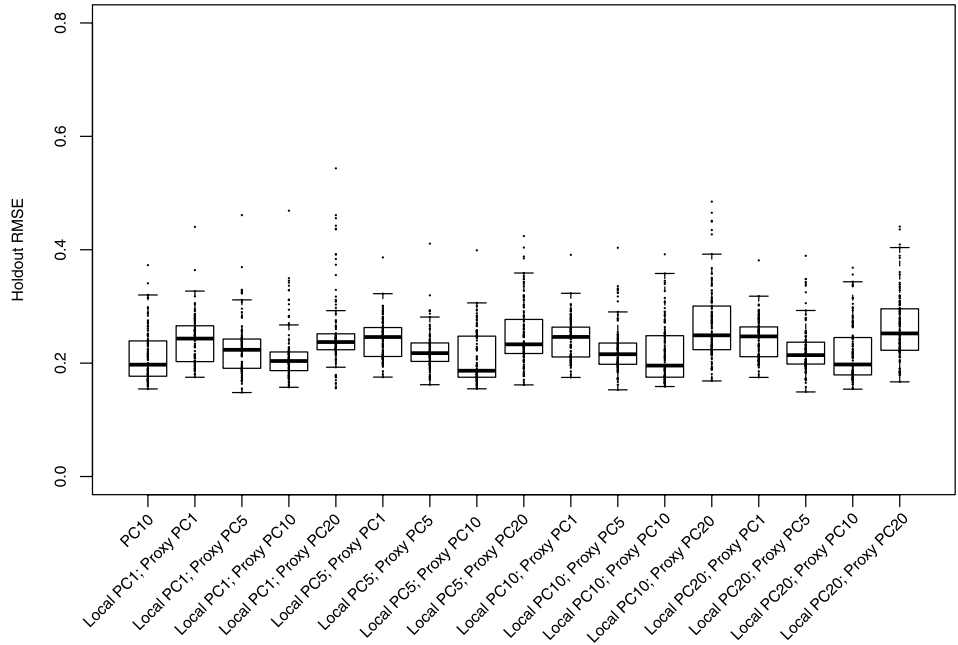}

\caption{Cross-validated RMSE on $30$-year holdout blocks for various
model specifications: regression on the first ten principal components
of the proxies and various two-stage models where global temperature is
regressed on principal components of local temperatures which are then
regressed on principal components of proxies.}\label{fig:alt3}
\end{figure}

As a final consideration, we employ a method similar to that used in
the original \citet{MaBrHu98} paper. This method takes account of the
fact that local proxies might be better predictors of local
temperatures than they are of global aggregate temperatures. For this
method, we again use the first $p$ principal components of the proxy
matrix but we also use the first $g$ principal components of the
$149\times1732$ local temperature matrix. We regress the CRU global
aggregate on the $g$ principal components of local temperature matrix,
and then we regress each of the $g$ local temperature principal
components on the $p$ proxy principal components. We can then use the
historical proxy principal components to backcast the local temperature
principal components thereby enabling us to backcast the global average
temperature.

We plot our results in Figure \ref{fig:alt3} and again include the
boxplot of ten principal components from Figure \ref{fig:alt1} for easy
reference. As before, there is simply not that much variation in holdout
RMSE across the various model specifications. No method is a clear winner.

%s4.2 ###
\subsection{Temperature reconstructions}

\begin{figure}

\includegraphics{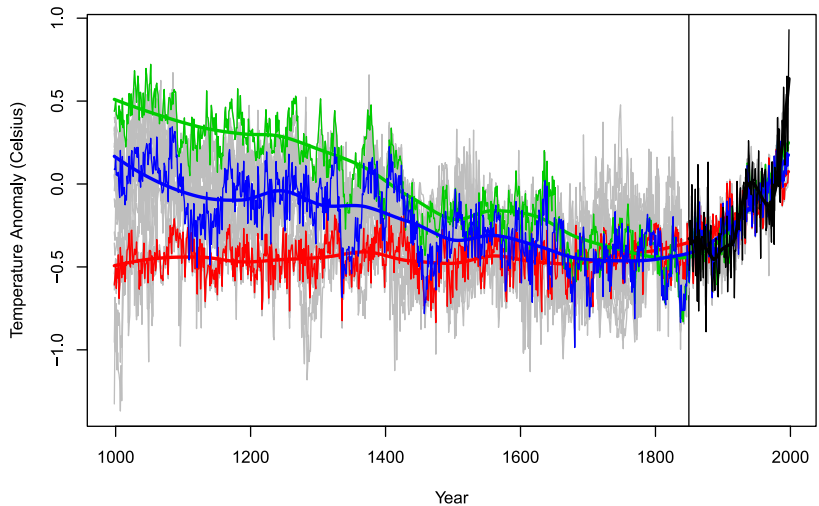}

\caption{Backcasts to $1000$ AD from the various models considered in
this section are plotted in gray. CRU Northern Hemisphere annual mean
land temperature is given by the thin black line with a smoothed
version given by the thick black line. Three forecasts are featured:
regression on one proxy principal component (red), regression on
ten proxy principal components (green), and the two-stage model
featuring five local temperature principal components and five proxy
principal components (blue).}\label{fig:altback}
\end{figure}

Each model discussed in Section \ref{altmodels} can form a historical
backcast. This backcast is simply the model's estimate $\hat{y}_k (
\mathbf{x}_t)$ of the Northern Hemisphere average temperature in a year
$t$ calculated by inputing the proxy covariates $\mathbf{x}_t$ in the same
year. The model index is $k$ which varies over all 27 models from
Section \ref{altmodels} (i.e., those featured in Figures \ref{fig:alt1}--\ref{fig:alt3}). We plot these
backcasts in Figure \ref{fig:altback} in gray and show the CRU average
in black. As can be seen, while these models all perform similarly in
terms of cross-validated RMSE, they have wildly different implications
about climate history.

%f14 ###

According to some of them (e.g., the ten proxy principal component
model given in green or the two-stage model featuring five local
temperature principal components and five proxy principal components
given in blue), the recent run-up in temperatures is not that abnormal,
and similarly high temperatures would have been seen over the last
millennium. Interestingly, the blue backcast seems to feature both a
Medieval Warm Period and a Little Ice Age whereas the green one shows
only increasing temperatures going back in time.

However, other backcasts (e.g., the single proxy principal component
regression featured in red) are in fact hockey sticks which correspond
quite well to backcasts such as those in \citet{MaBrHu99}. If they are
correct, modern temperatures are indeed comparatively quite alarming
since such temperatures are much warmer than what the backcasts
indicate was observed over the past millennium.

Figure \ref{fig:altback} reveals an important concern: models that
perform similarly at predicting the instrumental temperature series (as
revealed by Figures \ref{fig:alt1}--\ref{fig:alt3})
tell very different stories about the past. Thus, insofar as one judges
models by cross-validated predictive ability, one seems to have no
reason to prefer the red backcast in Figure \ref{fig:altback} to the
green even though the former %provokes alarm while the latter soothes.
suggests that recent temperatures are much warmer than those observed
over the past 1000 years while the latter suggests they are not.

A final point to note is that the backcasts plotted in Figure \ref
{fig:altback} are the raw backcasts themselves with no accounting for
backcast standard errors. In the next section, we take on the problem
of specifying a full probability model which will allow us to provide
accurate, pathwise standard errors.

%s5 ###
\section{Bayesian reconstruction and validation}\label{bayes}

%s5.1 ###
\subsection{Model specification}

In the previous section, we showed that a variety of different models
perform fairly similarly in terms of cross-validated RMSE while
producing very different temperature reconstructions. In this section,
we focus and expand on the model which uses the first ten principal
components of the proxy record to predict Northern Hemisphere CRU. We
chose this forecast for several reasons. First, it performed relatively
well compared to all of the others (see Figures \ref{fig:alt1}--\ref{fig:alt3}). Second, PC regression has a relatively
long history in the science of paleoclimatological reconstructions
[\citeauthor{MaBrHu98} (\citeyear{MaBrHu98}, \citeyear{MaBrHu99}), \citet{NRC06}]. Finally, when using OLS regression,
principal components up to and including the tenth were statistically
significant. While the \textit{t}-statistics and their associated \textit{p}-values themselves are uninterpretable due to the complex time series
and error structures, these traditional benchmarks can serve as
guideposts.

However, there is at least one serious problem with this model as it
stands: the residuals demonstrate significant autocorrelation not
captured by the autocorrelation in the proxies. Accordingly, we fit a
variety of autoregressive models to CRU time series. With an AR2 model,
the residuals showed very little autocorrelation.

So that we account for both parameter uncertainty as well as residual
uncertainty, we estimate our model using Bayesian procedures. Our
likelihood is given by
\begin{eqnarray*}
y_t &=& \beta_0 + \sum_{i=1}^{10} \beta_i x_{t,i} + \beta_{11} y_{t+1} +
\beta_{12} y_{t+2} + \epsilon_t, \\
\epsilon_t &\sim& N(0,\sigma^2).
\end{eqnarray*}
In our equation, $y_t$ represents the CRU Northern Hemisphere annual
land temperature in year $t$ and $x_{t,i}$ is the value of principal
component $i$ in year $t$. We note that the subscripts on the
right-hand side of the regression equation employ pluses rather than
the usual minuses because we are interested in backcasts rather than
forecasts. In addition to this, we use the very weakly informative priors
\begin{eqnarray*}
\vec\beta&\sim& N( \vec0, 1000 \cdot I ),
\\
\sigma&\sim& \operatorname{Unif}(0,100),
\end{eqnarray*}
where $\vec\beta$ is the 13 dimensional vector $(\beta_0, \beta
_1,\ldots, \beta_{12})^T$, $\vec0$ is a vector of 13 zeros, and $I$ is
the 13 dimensional identity matrix. This prior is sufficiently
noninformative that the posterior mean of $\vec\beta$ is, within
rounding error, equal to the maximum likelihood estimate. Furthermore,
the prior on $\sigma$ is effectively noninformative as $y_t$ is always
between $\pm1$ and therefore no posterior draw comes anywhere near the
boundary of $100$.

It is important to consider how our model accounts for the perils of
temperature reconstruction discussed above. First and foremost, we deal
with the problem of weak signal by building a simple model ($\operatorname{AR}2 + \mathrm{PC}10$)
in order to avoid overfitting. Our fully Bayesian model, which accounts
for parameter uncertainty, also helps attenuate some of the problems
caused by weak signal. Dimensionality reduction is dealt with via
principal components. PCs have two additional benefits. First, they are
well-studied in the climate science literature and are used in climate
scientists' reconstructions. Second, the orthogonality of principal
components will diminish the pernicious effects of spatial correlation
among the proxies. Finally, we address the temporal correlation of the
temperature series with the AR2 component of our model.

%s5.2 ###
\subsection{Comparison to other models}

An approach that is broadly similar to the above has recently appeared
in the climate literature [\citet{LiNycAmm07}] for purposes similar to
ours, namely, quantifying the uncertainty of a reconstruction. In fact,
\citet{LiNycAmm07} is highly unusual in the climate literature in that
its authors are primarily statisticians. Using a dataset of 14 proxies
from \citet{MaBrHu99}, \citet{LiNycAmm07} confirms the findings
of \citeauthor{MaBrHu98} (\citeyear{MaBrHu98}, \citeyear{MaBrHu99}) but attempts to take forecast error, parameter
uncertainty, and temporal correlation into account. They provide toy
data and code for their model here: %\footnote{In their Acknowledgments
%section \cite[]{LiNycAmm07}, the authors ``thank Dr. Michael Mann for
%providing the proxy data." It appears this data is unable to be
%shared, however, since the website linked to above contains the
%following note: ``The above two pseudo data sets are generated by
%adding white noise with unit variance to the standardized real data.
%They only serve as a toy data example to try the R code below.
%However, the results in Li et al. (Tellus, in press) are based on the
%real data instead of the pseudo data."}:
\url{http://www.image.ucar.edu/\textasciitilde boli/research.html}

Nevertheless, several important distinctions between their model and
ours exist. First, \citet{LiNycAmm07} make use of a dataset over ten
years old [\citet{MaBrHu99}] which contains only 14 proxies dating back
to 1000 AD and has instrumental records dating 1850--1980 AD. On the
other hand, we make use of the latest multi-proxy database [\citet{Mannetal08}] which contains 93 proxies dating back to 1000 AD and has
instrumental records dating 1850--1998 AD. Furthermore, \citet{LiNycAmm07} assume an AR2 structure on the errors from the model where
we assume the model is AR2 with covariates. Finally, and perhaps most
importantly, \citet{LiNycAmm07} estimate their model via generalized
least squares and therefore use (i) the parametric bootstrap in order
to account for parameter estimation uncertainty and (ii)
cross-validation to account overfitting the in-sample period (i.e., to
inflate their estimate of the error variance $\sigma$). On the other
hand, by estimating our model in a fully Bayesian fashion, we can
account for these within our probability model. Thus, our procedure can
be thought of as formalizing the approach of \citet{LiNycAmm07} and it
provides practically similar results when applied to the same set of
covariates (generalized least squares also produced practically
indistinguishable forecasts and backcasts though obviously narrower
standard errors).

%The same authors are currently working on a fully Bayesian model [\citet{LiNycAmm10}] which deserves mention.
At the time of this manuscript's submission, the same authors were working on a fully Bayesian
model which deserves mention [subsequently published as \citet{LiNycAmm10}].
In this paper, they integrate
data from three types of proxies measured at different timescales (tree
rings, boreholes, and pollen) as well as data from climate forcings
(solar irradiance, volcanism, and greenhouse gases) which are
considered to be external drivers of climate. Furthermore, they account
for autocorrelated error in both the proxies and forcings as well as
autocorrelation in the deviations of temperature from the model. While
the methodology and use of forcing data are certainly innovative, the
focus of \citet{LiNycAmm10} is not on reconstruction \textit{per se}; rather,
they are interested in validating their modeling approach taking as
``truth'' the output of a high-resolution state-of-the-art climate
simulation [\citet{Ametal07}]. Consequently, all data used in the paper
is synthetic and they concentrate on methodological issues,
``defer[ring] any reconstructions based on actual observations and
their geophysical interpretation to a subsequent paper'' [\citet{LiNycAmm10}].

Finally, \citeauthor{TinHuy10a} (\citeyear{TinHuy10a}, \citeyear{TinHuy10b}) have developed a hierarchical
Bayesian model to reconstruct the full temperature field. They fit the
model to experimental datasets formed by ``corrupting a number of the
[temperature] time series to mimic proxy observations'' [\citet{TinHuy10a}]. Using these datasets, they conduct what is in essence a
frequentist evaluation of their Bayesian model [\citet{TinHuy10a}] and
then compare its performance to that of the well-known RegEM algorithm
[\citet{TinHuy10b}]. Like \citet{LiNycAmm10}, however, they do not use
their model to produce temperature reconstructions from actual proxy
observations.

%s5.3 ###
\subsection{Model reconstruction}\label{recon}

We create a full temperature backcast by first initializing our model
with the CRU temperatures for 1999 AD and 2000 AD. We then perform a
``one-step-behind'' backcast, plugging these values along with the ten
principal component values for 1998 AD into the equation $y_t = \beta_0
+ \sum_{i=1}^{10} \beta_i x_{t,i} + \beta_{11} y_{t+1} + \beta_{12}
y_{t+2}$ to get a backcasted value for 1998 AD (using the posterior
mean of $\vec\beta$ as a plug-in estimator). Similarly, we use the CRU
temperature for 1999 AD, this backcasted value for 1998 AD, and the ten
principal component values for 1997 AD to get a backcasted value for
1997 AD. Finally, we then iterate this process one year at a time,
using the two most recent backcasted values as well as the current
principal component values, to get a backcast for each of the last
1000 years.

%f15 ###
\begin{figure}

\includegraphics{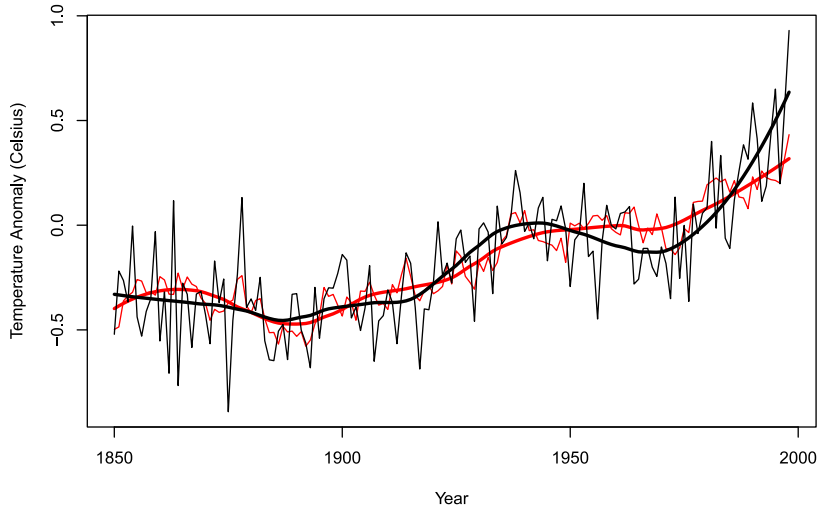}

\caption{In-sample Backcast from Bayesian Model of
Section \protect\ref{bayes}. CRU Northern Hemisphere annual mean land temperature is given
by the thin black line and a smoothed version is given by the thick
black line. The backcast is given by the thin red line and a smoothed
version is given by the thick red line. The model is fit on
1850--1998 AD.}\label{fig:bayes_backcast}
\end{figure}

We plot the in-sample portion of this backcast (1850--1998 AD) in
Figure \ref{fig:bayes_backcast}. Not surprisingly, the model tracks CRU
reasonably well because it is in-sample. However, despite the fact that
the backcast is both in-sample and initialized with the high true
temperatures from 1999 AD and 2000 AD, it still cannot capture either
the high level of or the sharp run-up in temperatures of the 1990s: it
is substantially biased low. That the model cannot capture run-up even
in-sample does not portend well for its ability to capture similar
levels and run-ups if they exist out-of-sample.

A benefit of our fully Bayesian model is that it allows us to assess
the error due to both (i) residual variance (i.e., $\epsilon_t$) and
(ii) parameter uncertainty. Furthermore, we can do this in a fully
pathwise fashion. To assess the error due to residual variance, we use
the one-step-behind backcasting procedure outlined above with two
exceptions. First, at each step, we draw an error from a $N(0,\sigma
^2)$ distribution and add it to our backcast. These errors then
propagate through the full path of backcast. Second, we perform the
backcast allowing $\sigma$ to vary over our samples from the posterior
distribution.

To assess the error due to the uncertainty in $\vec\beta$, we perform
the original one-step-behind backcast [i.e., without drawing an error
from the $N(0,\sigma^2)$ distribution]. However, rather than using the
posterior mean of $\vec\beta$, we perform the backcast for each of our
samples from the posterior distribution of~$\vec\beta$.

Finally, to get a sense of the full uncertainty in our backcast, we can
combine both of the methods outlined above. That is, for each draw from
the posterior of $\vec\beta$ and $\sigma$, we perform the
one-step-behind backcast drawing errors from the $N(0,\sigma^2)$
distribution. This gives one curve for each posterior draw, each
representing a draw of the full temperature series conditional on the
data and the model. Taken together, they form an approximation to the
full posterior distribution of the temperature series.

%f16 ###
\begin{figure}

\includegraphics{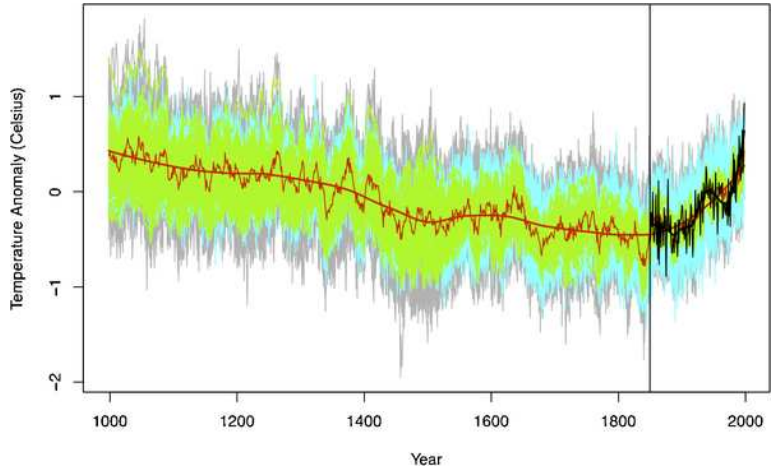}

\caption{Backcast from Bayesian Model of Section \protect\ref{bayes}. CRU
Northern Hemisphere annual mean land temperature is given by the thin
black line and a smoothed version is given by the thick black line. The
forecast is given by the thin red line and a smoothed version is given
by the thick red line. The model is fit on 1850--1998 AD and
backcasts 998--1849 AD. The cyan region indicates uncertainty due
to $\epsilon_t$, the green region indicates uncertainty due to $\vec
\beta$, and the gray region indicates total uncertainty.}\label
{fig:bayes_backcast_error}
\end{figure}

We decompose the uncertainty of our model's backcast by plotting the
curves drawn using each of the methods outlined in the previous three
paragraphs in Figure \ref{fig:bayes_backcast_error}. As can be seen, in
the modern instrumental period the residual variance (in cyan)
dominates the uncertainty in the backcast. However, the variance due to
$\vec\beta$ uncertainty (in green) propagates through time and becomes
the dominant portion of the overall error for earlier periods. The
primary conclusion is that failure to account for parameter uncertainty
results in overly confident model predictions.

As far as we can tell, no effort at paleoclimatological global
temperature reconstruction of the past 1000 years has used a fully
Bayesian probability model to incorporate parameter uncertainty into
the backcast estimates [in fact, the aforementioned \citet{LiNycAmm07}
paper is the only paper we know of that even begins to account for
uncertainty in some of the parameters; see \citet{Hasetal06} for a
Bayesian model used for reconstructing the local prehistoric climate in
Glendalough, Ireland]. The widely used approach in the climate
literature is to estimate uncertainty using residuals (usually from a
holdout period). Climate scientist generally report less accurate
reconstructions in more distant time periods, but this is due to the
fact that there are fewer proxies that extend further back into time
and therefore larger validation residuals.

%f17 ###
\begin{figure}

\includegraphics{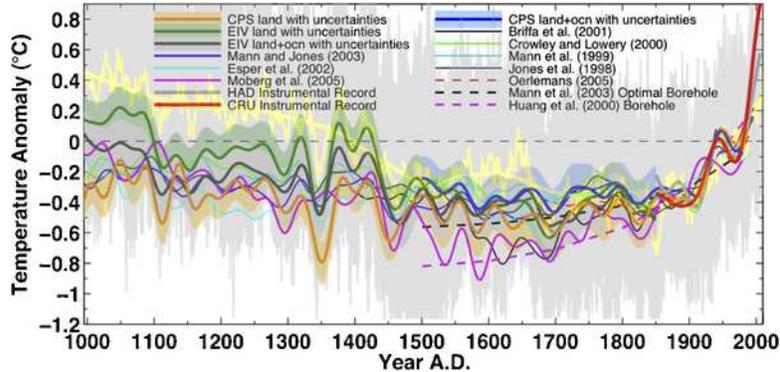}

\caption{This figure modifies Figure $3$ from Mann et~al. (\protect\citeyear{Mannetal08}). We
take that figure and superimpose the backcast from Bayesian model of
Section \protect\ref{bayes}. The backcast is given by the thin yellow line,
the smoothed backcast by a thick yellow line, and the backcast error in
gray.}\label{fig:allbackcast}
\end{figure}

%s5.4 ###
\subsection{Comparison to other reconstructions and posterior calculations}

\hspace{-5pt}What is most interesting is comparing our backcast to those from
\citet{Mannetal08} as done in Figure \ref{fig:allbackcast}. We see that our
model gives a backcast which is very similar to those in the
literature, particularly from 1300 AD to the present. In fact, our
backcast very closely traces the \citet{Mannetal08} EIV land backcast,
considered by climate scientists to be among the most skilled. Though
our model provides slightly warmer backcasts for the years 1000--1300
AD, we note it falls within or just outside the uncertainty bands of
the \citet{Mannetal08} EIV land backcast even in that period. Hence, our
backcast matches their backcasts reasonably well.

The major difference between our model and those of climate scientists,
however, can be seen in the \textit{large width} of our uncertainty bands.
Because they are pathwise and account for the uncertainty in the
parameters (as outlined in Section~\ref{recon}), they are much larger
than those provided by climate scientists. In fact, our uncertainty
bands are so wide that they \textit{envelop} all of the other backcasts in
the literature. Given their ample width, it is difficult to say that
recent warming is an extraordinary event compared to the last 1000
years. For example, according to our uncertainty bands, it is possible
that it was as warm in the year 1200 AD as it is today. In contrast,
the reconstructions produced in \citet{Mannetal08} are completely pointwise.

Another advantage of our method is that it allows us to calculate
posterior probabilities of various scenarios of interest by simulation
of alternative sample paths. For example, 1998 is generally considered
to be the warmest year on record in the Northern Hemisphere. Using our
model, we calculate that there is a 36\% posterior probability that
1998 was the warmest year over the past thousand. If we consider
rolling decades, 1997--2006 is the warmest on record; our model gives
an 80\% chance that it was the warmest in the past 1000 years.
Finally, if we look at rolling 30-year blocks, the posterior
probability that the last 30 years (again, the warmest on record) were
the warmest over the past thousand is 38\%.

Similarly, we can look at posterior probabilities of the run-up in (or
derivative of) temperatures in addition to the levels. For this
purpose, we defined the ``derivative'' as the difference between the
value of the loess smooth of the temperature series (or reconstruction
series) in year $t$ and year $t-k$. For $k=10$, $k=30$, and $k=60$, we
estimate a zero posterior probability that the past 1000 years
contained run-ups larger than those we have experienced over the past
ten, 30, and 60 years (again, the largest such run-ups on record). This
suggests that the temperature derivatives encountered over recent
history are unprecedented in the millennium. While this does seem
alarming, we should temper our alarm somewhat by considering again
Figure \ref{fig:bayes_backcast} and the fact that the proxies seem
unable to capture the sharp run-up in temperature of the 1990s. That
is, our posterior probabilities are based on derivatives from our
model's proxy-based reconstructions and we are comparing these
derivatives to derivatives of the \textit{actual} temperature series;
insofar as the proxies cannot capture sharp run-ups, our model's
reconstructions will not be able to either and therefore will tend to
understate the probability of such run-ups.

%s5.5 ###
\subsection{Model validation}

Though our model gives forecasts and backcasts that are broadly
comparable to those provided by climate scientists, our approach
suggests that there is substantial uncertainty about the ability of the
model to fit and predict new data. Climate scientists estimate
out-of-sample uncertainty using only two holdout blocks: one at the
beginning of the instrumental period and one at the end. We pursue that
strategy here. First, we fit on 1880--1998 AD and attempt to backcast
1850--1879 AD. Then, we fit on 1850--1968 AD and forecast 1969--1998
AD. These blocks are arguably the most interesting and important
because they are not ``tied'' at two endpoints. Thus, they genuinely
reflect the most important modeling task: reconstruction.

%f18 ###
\begin{figure}

\includegraphics{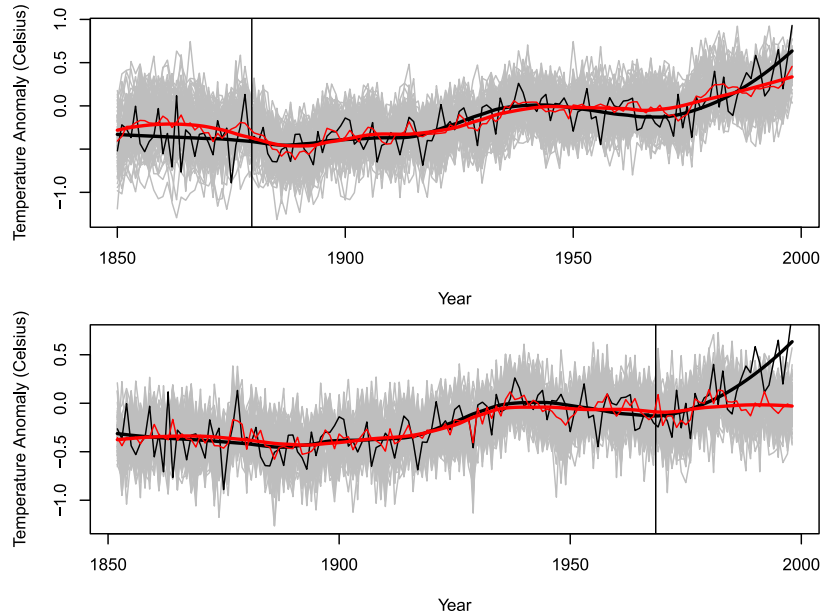}

\caption{Predictions from the Bayesian model of Section \protect\ref{bayes}
when the first $30$ years of instrumental data are held out \textup{(top)} and
when the last $30$ years of instrumental data are held out \textup{(bottom)}. CRU
is given in black and the model predictions in red. The raw data and
predictions are given by the thin lines and loess smooths are given by
the thick lines. Uncertainty bands are indicated by the gray
region.}\label{fig:bayes_firstlast}
\end{figure}

Figure \ref{fig:bayes_firstlast} illustrates that the model seems to
perform reasonably well on the first holdout block. Our reconstruction
regresses partly back toward the in-sample mean. Compared to the actual
temperature series, it is biased a bit upward. On the other hand, the
model is far more inaccurate on the second holdout block, the modern
period. Our reconstruction, happily, does not move toward the in-sample
mean and even rises substantively at first. Still, it seems there is
simply not enough signal in the proxies to detect either the high
levels of or the sharp run-up in temperature seen in the 1990s. This is
disturbing: if a model cannot predict the occurrence of a sharp run-up
in an out-of-sample block which is contiguous with the in-sample
training set, then it seems \textit{highly unlikely} that it has power to
detect such levels or run-ups in the more distant past. It is even more
discouraging when one recalls Figure \ref{fig:bayes_backcast}: the
model cannot capture the sharp run-up even \textit{in-sample}. In sum,
these results suggest that the 93 sequences that comprise the
1000-year-old proxy record simply lack power to detect a sharp
increase in temperature.\footnote{On the other hand, perhaps our model
is unable to detect the high level of and sharp run-up in recent
temperatures because anthropogenic factors have, for example, caused a
regime change in the relation between temperatures and proxies. While
this is certainly a consistent line of reasoning, it is also fraught
with peril for, once one admits the possibility of regime changes in
the instrumental period, it raises the question of whether such changes
exist elsewhere over the past 1000 years. Furthermore, it implies that
up to half of the already short instrumental record is corrupted by
anthropogenic factors, thus undermining paleoclimatology as a
statistical enterprise.}

As mentioned earlier, scientists have collected a large body of
evidence which suggests that there was a Medieval Warm Period (MWP) at
least in portions of the Northern Hemisphere. The MWP is believed to
have occurred c. 800--1300 AD (it was followed by the Little Ice
Age). It is widely hoped that multi-proxy models have the power to
detect (i) how warm the Medieval Warm Period was, (ii) how sharply
temperatures increased during it, and (iii) to compare these two
features to the past decade's high temperatures and sharp run-up. Since
our model cannot detect the recent temperature change, detection of
dramatic changes hundreds of years ago seems out of the question.

This is not to say that the proxy record is unrelated to temperatures.
We can compare our model's RMSE in these two holdout periods to various
null models which we know have no signal. That is, we can perform a
test similar to that of Section \ref{interpextrap}. On each holdout
block, we generate a $149 \times 93$ matrix of pseudo-proxies from each of
the six null models known to be independent of the temperature series.
Then, analogously to our model, we take the first ten principal
components of these pseudo-proxies, regress the in-sample temperature
on the ten in-sample principal components, and compute the RMSE on the
holdout block. We perform this procedure 1000 times for each holdout block and then calculate
the percentage of time that the model fit to the pseudo-proxies beats
our model.

%t2 ###
\begin{table}
\tablewidth=290pt
\caption{Percent of time various null models outperform the Bayesian
model of Section \protect\ref{bayes}}\label{pval}
\begin{tabular*}{\tablewidth}{@{\extracolsep{\fill}}lcc@{}}
\hline
\textbf{Pseudo-proxy} & \textbf{First block $\bolds{p}$-value} & \textbf{Last block
$\bolds{p}$-value} \\
\hline
White Noise & \hphantom{0}0.0\% & \hphantom{0}0.0\% \\
AR1(0.25) & \hphantom{0}0.1\% & \hphantom{0}0.0\% \\
AR1(0.4) & \hphantom{0}0.1\% & \hphantom{0}0.0\% \\
Empirical AR1 & 24.1\% & 20.6\% \\
Brownian Motion & 16.4\% & 32.2\% \\
%Pseudo-Bridge & 21.1\% & 2.6\% \\
\hline
\end{tabular*}
\end{table}

Our model, with an RMSE of 0.26 on the first holdout block and an RMSE
of 0.36 on the second handily outperforms the relatively
unsophisticated white noise and weak AR1 process pseudo-proxies (see
Table \ref{pval}). Again, this is not surprising. These pseudo-proxies
cannot capture the local dependence in the instrumental record, so they
regress sharply to the in-sample mean. On the other hand, the Empirical
AR1 processes and Brownian motion have more complex local structure so
they provide respectable competition to our model. These models capture
only the local dependence in the temperature record: in the long term,
forecasts based off the AR1 processes will slide slowly back to the
in-sample mean and forecasts based off Brownian motion will wander
aimlessly. Taken together, it follows that our model is at best weakly
significant relative to the Empirical AR1 process or Brownian motion on
either holdout block.

In tandem, Figure \ref{fig:bayes_firstlast} and Table \ref{pval}
should make us very cautious about using our model to extrapolate, even
with wide standard errors. The second panel of Figure \ref
{fig:bayes_firstlast} demonstrates that these standard errors are too
narrow even for very temporally short forecasts. While we are able to
replicate the significance tests in \citet{MaBrHu98}, our Table \ref
{pval} shows that our model does not pass ``statistical significance''
thresholds against savvy null models. Ultimately, what these tests
essentially show is that the 1000-year-old proxy record has little
power given the limited temperature record.

%s6 ###
\section{Conclusion}

Research on multi-proxy temperature reconstructions of the earth's
temperature is now entering its second decade. While the literature is
large, there has been very little collaboration with university-level,
professional statisticians [\citet{WegScoSaid06}, \citet{Wegman06}]. Our paper is
an effort to apply some modern statistical methods to these problems.
While our results agree with the climate scientists findings in some
respects, our methods of estimating model uncertainty and accuracy are
in sharp disagreement. %Consequently, both climate change ``alarmists"
%and ``skeptics" will find encouragement in our work.

On the one hand, we conclude unequivocally that the evidence for a
``long-handled'' hockey stick (where the shaft of the hockey stick
extends to the year 1000 AD) is lacking in the data. The fundamental
problem is that there is a limited amount of proxy data which dates
back to 1000 AD; what is available is weakly predictive of global
annual temperature. Our backcasting methods, which track quite closely
the methods applied most recently in \citet{Mann08} to the same data,
are unable to catch the sharp run up in temperatures recorded in the
1990s, even in-sample. As can be seen in Figure \ref
{fig:bayes_backcast}, our estimate of the run up in temperature in the
1990s has a much smaller slope than the actual temperature series.
Furthermore, the lower frame of Figure \ref{fig:bayes_firstlast}
clearly reveals that the proxy model is not at all able to track the
high gradient segment. Consequently, the long flat handle of the hockey
stick is best understood to be a feature of regression and less a
reflection of our knowledge of the truth. Nevertheless, the
temperatures of the last few decades have been relatively warm compared
to many of the 1000-year temperature curves sampled from the posterior
distribution of our model.

Our main contribution is our efforts to seriously grapple with the
uncertainty involved in paleoclimatological reconstructions. Regression
of high-dimensional time series is always a complex problem with many
traps. In our case, the particular challenges include (i) a short
sequence of training data, (ii) more predictors than observations,
(iii) a very weak signal, and (iv) response and predictor variables
which are both strongly autocorrelated. The final point is particularly
troublesome: since the data is not easily modeled by a simple
autoregressive process, it follows that the number of truly independent
observations (i.e., the effective sample size) may be just too small
for accurate reconstruction.

Climate scientists have greatly underestimated the uncertainty of
proxy-based reconstructions and hence have been overconfident in their
models. We have shown that time dependence in the temperature series is
sufficiently strong to permit complex sequences of random numbers to
forecast out-of-sample reasonably well fairly frequently (see
Figures \ref{fig:boxplot} and \ref{fig:rmseyr}). Furthermore, even proxy-based models with
approximately the same amount of reconstructive skill (Figures \ref
{fig:alt1}--\ref{fig:alt3}), produce strikingly
dissimilar historical backcasts (Figure \ref{fig:altback}); some of these look like hockey sticks
but most do not.

Natural climate variability is not well understood and is probably
quite large. It is not clear that the proxies currently used to predict
temperature are even predictive of it at the scale of several decades
let alone over many centuries. Nonetheless, paleoclimatoligical
reconstructions constitute only one source of evidence in the AGW debate.

Our work stands entirely on the shoulders of those environmental
scientists who labored untold years to assemble the vast network of
natural proxies. Although we assume the reliability of their data for
our purposes here, there still remains a considerable number of
outstanding questions that can only be answered with a free and open
inquiry and a great deal of replication.

\section*{Acknowledgments}
We thank Editor Michael Stein, two anonymous referees, and Tilmann
Gneiting for their helpful suggestions on our manuscript. We also thank
our colleagues Larry Brown and Dean Foster for many helpful conversations.

\begin{supplement}%[id-suppA]
\stitle{Code repository for ``A statistical analysis of multiple
temperature proxies: Are reconstructions of surface temperatures over
the last 1000 years reliable?''}
\slink[doi]{10.1214/10-AOAS398SUPP}
\slink[url]{http://lib.stat.cmu.edu/aoas/398/supplement.zip}
\sdatatype{.zip}
\sdescription{This repository archives all data and code used for ``A
statistical analysis of multiple temperature proxies: Are
reconstructions of surface temperatures over the last 1000 years
reliable?'' In particular, it contains code to make all figures and
tables featured in the paper.}
\end{supplement}

%suskaldyti doi

% imsref loaded by dianan, 2011-01-19 15:29:50
% imsref loaded by dianan, 2011-01-20 10:44:17
%
% imsref loaded by dianan, 2011-01-20 11:30:09

\printaddresses

\end{document}